\documentclass[useAMS,usenatbib]{mn2e}
\usepackage{graphicx}
\usepackage{amsmath,amssymb}
\usepackage{multirow}
\def\etal{{\it et al.}}
\def\KFL{{Kwan et al.}}
\def\LCDM{$\Lambda$CDM}
\def\PRL#1{Phys.\ Rev.\ Lett.\ {\bf#1}} \def\PR#1{Phys.\ Rev.\ {\bf#1}}
\def\ApJ#1{ApJ {\bf#1}} \def\AJ#1{AJ {\bf#1}}
\def\AaA#1{Astron.\ Astrophys.\ {\bf#1}} 
\def\MNRAS#1{MNRAS {\bf#1}}
\def\CQG#1{Class.\ Quantum Grav.\ {\bf#1}}
\def\GRG#1{Gen.\ Relativ.\ Grav.\ {\bf#1}}
\def\lsim{\mathop{\hbox{${\lower3.8pt\hbox{$<$}}\atop{\raise0.2pt\hbox{$
\sim$}}$}}} \def\ns#1{_{\rm #1}}
\def\gsim{\mathop{\hbox{${\lower3.8pt\hbox{$>$}}\atop{\raise0.2pt\hbox{$
\sim$}}$}}}\def\goesas{\mathop{\sim}\limits} 
\def\Z#1{_{\lower2pt\hbox{$\scriptstyle#1$}}}
\def\X#1{_{\lower2pt\hbox{$\scriptstyle#1$}}}
\def\w#1{\,\hbox{#1}} \def\kmsMpc{\w{km}\;\w{sec}^{-1}\w{Mpc}^{-1}}
\def\de{\delta}\def\rh{\rho} \def\Hm{H\Z0} \def\Hb{\bar H\Z0}
\def\OM{\Omega} \def\bOM{\bar\OM} \def\OmMn{\OM_{\rmn{m}0}}
\def\finfty{{\mathop{\hbox{\it fi}}}}\def\Fi{\hbox{\footnotesize\it fi}}
\def\ta{\tau} \def\tv{\ta_{\rmn{v}}} \def\tw{\ta}
\def\ab{{\bar{a}}} \def\av{a_{\rmn{v}}} \def\aw{a_{\rmn{w}}}
\def\Hv{H_{\rmn{v}}} \def\Hw{H_{\rmn{w}}} \def\hr{h_r} \def\ti{t_{\rmn{i}}}
\def\etw{\eta_{\rmn{w}}} \def\etb{\bar\eta} \def\dL{d\Z L}
\def\gb{\bar\gamma} \def\rw{r_{\rmn{w}}}
\def\fw{f_{\rmn{w}}} \def\fwi{f_{\rmn{wi}}} \def\FF{{\cal F}}
\def\fv{f_{\rmn{v}}} \def\fvi{f_{\rmn{vi}}} \def\fvf{\left(1-\fv\right)}
\def\fvn{f_{\rmn{v}0}} \def\bH{\bar H} \def\QQ{{\cal Q}} \def\tb{t'}
\def\OMM{\bOM_{\rmn{m}}}\def\OMk{\bOM_{\rmn{k}}}\def\OMQ{\bOM_{\QQ}}
\def\pt{\partial}\def\rhb{\bar\rh}\def\kv{k_{\rmn{v}}} \def\Hh{H}
\def\Der#1#2{{#1\hphantom{#2}\over#1#2}}
\def\Dts{\mathop{\hbox{$\Der\dd t$}}}
\def\Dtc{\mathop{\hbox{$\Der\dd\tw$}}}
\def\RV{R\Z V} \def\th{\theta} \def\si{\sigma}
\def\cosfit{\texttt{simple$\_$cosfitter}}
\def\scrM{\mathcal{M}} \def\zmin{z_{\rm min}}
\def\sne{SNe Ia}
\def\sn{SN Ia}
\def\bq{\begin{quotation}}
\def\eq{\end{quotation}}
\def\beq{\begin{equation}}
\def\eeq{\end{equation}}
\def\bea{\begin{eqnarray}}
\def\eea{\end{eqnarray}}
\def\dd{\rmn{d}}

\def\Omm{$\OmMn$}
\def\kmsmpc{km s$^{-1}$ Mpc$^{-1}~$}
\def\H=#1{$\Hm = {#1}$~\kmsmpc}
\def\Om=#1{$\OM\Z{M0} = {#1}$}
\def\c=#1{$\chi^2 = {#1}$}
\catcode`@=11 \def\@volume{\rm, accepted}\def\@pagerange{}\def\@pubyear{2010}
\catcode`@=12

\title[Supernova tests of the timescape cosmology]{Supernova tests of the
timescape cosmology}
\author[Smale \& Wiltshire]{Peter R. Smale\thanks{E-mail:
peter.smale@pg.canterbury.ac.nz} \& David L. Wiltshire\thanks{E-mail:
David.Wiltshire@canterbury.ac.nz}\\
Department of Physics \& Astronomy, University of Canterbury,
Private Bag 4800, Christchurch 8140, New Zealand\\
}
\date{Received 30 September 2010; in final form 26 November 2010}
\begin{document}
\maketitle
\label{firstpage}
\begin{abstract}
The timescape cosmology has been proposed as a viable alternative to
homogeneous cosmologies with dark energy. It realises cosmic acceleration
as an apparent effect that arises in calibrating average cosmological
parameters in the presence of spatial curvature and gravitational
energy gradients that grow large with the growth of inhomogeneities at late
epochs. Recently Kwan, Francis and Lewis have claimed that the timescape
model provides a relatively poor fit to the Union and Constitution supernovae
compilations, as compared to the standard \LCDM\ model. We show this
conclusion is a result of systematic issues in supernova light curve fitting,
and of failing to exclude data below the scale of statistical homogeneity,
$z\lsim0.033$. Using all currently available supernova datasets (Gold07,
Union, Constitution, MLCS17, MLCS31, SDSS-II, CSP, Union2),
and making cuts at the statistical homogeneity scale, we show that data
reduced by the SALT/SALT-II fitters provides Bayesian evidence that favours
the spatially flat \LCDM\ model over the timescape model, whereas data reduced
with MLCS2k2 fitters gives Bayesian evidence which favours the timescape
model over the \LCDM\ model. We discuss the questions of
extinction and reddening by dust, and of intrinsic colour variations in
supernovae which do not correlate with the decay time, and the likely
impact these systematics would have in a scenario consistent with the
timescape model.
\end{abstract}

\section{Introduction}

The enigma of a late epoch apparent acceleration of the expansion of the
universe is one of the greatest fundamental challenges we have ever faced
in theoretical cosmology. Our standard understanding is that
this is probably due to the present--day universe being dominated by a
cosmological constant or other fluid--like ``dark energy'' with an equation
of state, $P=w\rho$, which violates the strong energy condition. However,
this conclusion is based on the assumption that the universe is
well--described by a geometry which is exactly a homogeneous and isotropic
Friedmann--Lema\^{\i}tre--Robertson--Walker (FLRW) model,
with additional Newtonian perturbations. Observationally, however, the
assumption of homogeneity is open to challenges.

At the epoch of last scattering the matter distribution was certainly
very homogeneous, given the evidence of the cosmic microwave background
(CMB) radiation. Furthermore, if we look back to galaxies at large
redshifts early in cosmic history, such as those in the Hubble Deep Field,
then the distribution is relatively homogeneous.
However, in the intervening aeons the matter distribution has
become very inhomogeneous through the growth of structure. Large scale
surveys reveal the present epoch universe to possess a cosmic web of
structures, dominated in volume by voids, with galaxy clusters strung in
sheets that surround the voids, and filaments that thread them. Surveys
indicate that 40-50\% of the nearby universe is in voids of a characteristic
diameter of $30h^{-1}$ Mpc \citep{HV1,HV2}. Once numerous additional
minivoids \citep{minivoids} are included, the present epoch universe
appears to be void--dominated. Statistical homogeneity of this structure
appears only to be reached by averaging on scales of order at least
$100h^{-1}$ Mpc or more, where $h$ is the dimensionless
parameter related to the Hubble constant by $H\Z0=100h\kmsMpc$. The problem
of fitting a smooth geometry to a universe with such a lumpy matter
distribution \citep{fit1,fit2} is a nontrivial one, but central to
relating observations to the numerical values of the averaged parameters
which describe the Universe and its evolution as a whole.

The timescape (TS) cosmological model \citep{clocks,sol,obs}
has been proposed as a potentially viable alternative to homogeneous
cosmology with fluid-like dark energy. It begins from the premise,
consistent with observations of void statistics, that the present epoch
universe is strongly inhomogeneous on scales below the Baryon
Acoustic Oscillation (BAO) scale of order $100h^{-1}$ Mpc, while exhibiting
a variance of density of order 8\% in density for sample volumes larger
than this scale \citep{Hogg05,SylosLabini09}. This is consistent with the
growth of density contrasts with initially small fluctuations $\de\rh/\rh
\goesas10^{-4}$ in dark matter at last scattering, without assuming evolution
by a single Friedmann scale factor for the whole universe \citep{obs}.
The reason for a $100h^{-1}$ Mpc cutoff for this ``scale of statistical
homogeneity'' is that below this scale density contrasts at last scattering
are amplified by the acoustic waves in the primordial plasma.

In the two--scale model of \citet{clocks,sol,obs} cosmic evolution
is determined by a Buchert average \citep{Buchert00} over spatially flat wall
regions, assumed to contain all gravitationally bound structures, and
negatively curved voids. In the timescape scenario there is a crucial
ingredient in the physical interpretation of the Buchert average, which
has not been pursued in other investigations using the Buchert
formalism. (See, e.g., \citet{Buchert08} for a recent review.)
In particular, given that the Buchert average is a statistical average
taken by volume, then for a universe which is void--dominated at late
epochs the local geometry at a volume--average location can differ
considerably from that of sources and observers in gravitationally bound
systems within galaxies within the wall regions. It is
hypothesised that quasi-local gravitational energy gradients result
from the density and spatial curvature gradients between the two locations.
In particular, a substantial difference in clock rates arises
cumulatively over the lifetime of the universe from the tiny
relative volume deceleration of voids as compared to wall regions
\citep{equiv}. The magnitude of this relative deceleration, typically of order
$10^{-10}\,$m$\,$s$^{-2}$ over much of the lifetime of the universe,
is consistent with weak--field expectations, and may be understood
conceptually in terms of a generalisation of the equivalence principle
\citep{equiv}.

The faintness of type Ia supernovae (SNe Ia) relative to
the expectations of matter--dominated homogeneous cosmologies
then arises both as a consequence of: (i) the changes to cosmic evolution that
arise in a Buchert average of the Einstein equations; and (ii) the growing
differences in the calibration of the notional clocks and rulers of an ideal
volume--average observer as compared to the actual clocks and rulers of
observers confined to gravitationally bound systems. In particular, cosmic
acceleration is an apparent effect: a volume--average observer in a void
would not infer any cosmic acceleration, but observers in galaxies do
\citep{clocks}. This provides a natural solution to the cosmic coincidence
problem, since apparent acceleration is only registered once the void
fraction reaches a critical threshold of 59\%, typically near a redshift
$z\goesas0.9$ \citep{clocks}.

\citet*{LNW} considered some preliminary quantitative tests of the TS
model, concentrating in particular on the fit of the 182 \sn\ distance
moduli in the Gold dataset of \citet{Riess06} (Riess07).
It was found that the TS model gave a relatively good fit to the Riess07
data, as compared to the spatially flat \LCDM\ model. This conclusion
has recently been challenged by \citet*{KFL}, who concluded that the TS
model proved to be a poorer fit than the spatially flat \LCDM\ model when
tested with the more recent and larger Union \citep{Kowalski08} and
Constitution \citep{Hicken09} \sn\ datasets. Furthermore, the best--fit
value of the TS dressed matter density parameter was found by \citet{KFL}
to be driven to values an order of magnitude smaller than the \citet{LNW}
fit $\OmMn=0.33^{+0.11}_{-0.16}$.

In this paper, we reply to \citet{KFL} by investigating the fit of the
TS model to a number of recent SneIa datasets fit by different data
reduction methods, giving careful consideration to the systematic issues
that arise when one is dealing with a nonstandard cosmological model. These
issues were not considered by \KFL{} In the course of our investigations we
have uncovered one error in the \citet{LNW} paper: on account of a bug in a
numerical code the value of the Bayes factor that was quoted there was
incorrect\footnote{This numerical error in the routine used in the
Bayesian integration does not affect any other numerical results
given in \citet{LNW}.}.
Rather than a Bayes factor, $\ln B=0.27$ in favour of the TS model
over the spatially flat \LCDM\ model, the Bayes factor relative to
the Riess07 data is in fact $\ln B=-1.12$ for the given priors with
$0.01\le\OmMn\le0.5$. In other
words, rather than being statistically indistinguishable, the Riess07 data
in fact provides mild positive evidence in favour of the spatially flat
\LCDM\ model as compared to the TS model. This numerical
error was not raised by \citet{KFL}, however. Rather, the main
criticism\footnote{In addition to the principal discussion of fits to SneIa
data, \citet{KFL} also make some apparently critical remarks about the
TS model which deserve some reply. Firstly, \KFL\ comment that Birkhoff's
theorem is not relevant to the TS cosmology. However, Birkhoff's theorem --
namely the statement that the unique spherically symmetric solution
of the {\em vacuum} Einstein equations is the Schwarzschild solution
-- is not relevant in any circumstance in which the energy--momentum
tensor of matter is non-zero, which also includes the standard FLRW cosmology
with a dust or perfect fluid source. This comment is therefore gratuitous.
Secondly, \KFL\ also state that the Newtonian limit is not relevant
in the TS cosmology. This remark is simply incorrect, and is presumably
based on incomplete reading of \citet{clocks,equiv}. In the absence of
a simple background geometry, which is the case in a genuinely
inhomogeneous background, then the correct derivation of the Newtonian
limit, together with post--Newtonian corrections, is a subtle and nontrivial
problem which is much more complicated than the case of the standard
perturbed FLRW universe, as was discussed in Sec.~8.4 of \citet{clocks}.
One reason for proposing an extension of the
strong equivalence principle to the cosmological equivalence principle
\citep{equiv} is the hope that it might ultimately provide a framework for
establishing a Newtonian limit in the case that there are strong
inhomogeneities in the background geometry. The fact that a
post--Newtonian framework has not yet been worked out in detail is simply
due to the TS cosmology being work in progress. Finally, the claim that
the TS model suffers a ``failure to provide any predictions on cosmological
parameters that are directly observable'' \citep{KFL} is refuted by
\citet{obs}, where several cosmological tests with the potential to
distinguish the TS cosmology from the standard FLRW cosmology are discussed
in detail.}
of \citet{KFL} is a relatively poor fit of the TS model to the Union and
Constitution datasets. As we will see, this conclusion relies on
the na\"{\i}ve use of data which has been reduced assuming the standard
cosmology by the SALT methods. One particularly important
consideration in analysing the data is that a cut should always be made at the
scale of statistical homogeneity expected in the TS model -- a fact that
\KFL\ omitted. We shall see that depending on the dataset and fitting
method used, in some cases the TS model provides a fit that is either
statistically indistinguishable from \LCDM, or mildly favoured or disfavoured
on Bayesian evidence. However, given
other unknown systematic differences between the variants of the SALT and
MLCS methods, at present it is fair to say one cannot reliably distinguish the
TS and \LCDM\ models on the basis of \sne\ data alone.

The outline of the paper is as follows.
In section~\ref{sect:ts} we give a brief overview of the timescape cosmology,
while in section~\ref{sect:fitters} we given an overview of the \sn\ light
curve data reduction methods. In section~\ref{sect:results} we
analyse the currently available data, considering in particular the issues of
recalibrating the SALT light curve fitter, exclusion of data below the
statistical homogeneity scale, and differences between the various \sne\
datasets. In section~\ref{systematics} we discuss the impact that unknown
systematic issues, particularly concerning intrinsic colours variations and
reddening and extinction, may have on the comparison of the TS model
with the standard cosmology.

\section[]{The timescape model} \label{sect:ts}
In keeping with current observations that the volume of the present epoch
universe is dominated by voids (typically of diameter $\sim30h^{-1}$ Mpc
and smaller) which are separated and threaded by walls and filaments
containing clusters of galaxies, the timescape model is based on
two scales: the spatially flat wall regions which contain gravitationally
bound structures, and the voids, which are negatively curved. Large gradients
in Ricci scalar curvature are assumed to exist between the walls/filaments
and the voids, consistent with observations that the latter have density
contrasts close to $\de\rho/\rho\goesas-1$ \citep{HV1,HV2}. At any location
the local spatial curvature is in general different to any global average,
and in order to make sense of average cosmological parameters this variance
in local geometries must also be taken into account \citep{clocks}.
Observers in any region who try to fit a single FLRW geometry based on the
mistaken assumption that the global spatial curvature is the
same as the local value will determine different cosmological parameters.

As observers in galaxies, our local average geometry, up to a scale
enclosing gravitationally bound galaxy clusters, is assumed to be
spatially flat on average and marginally expanding at the boundary,
with a FLRW--type geometry with scale factor $\aw$,
\beq\label{eq:wallmetric}
\dd s^2\Z{\Fi}=-\dd\tw^2+\aw^2(\tw)[\dd\etw^2+\etw^2\dd\OM^2].
\eeq
Finite infinity\footnote{\citet{fit1} gave a qualitative definition of
finite infinity. The definition adopted here \citep{clocks} is one possible
realisation.} \citep{fit1}, denoted $\finfty$, demarcates the boundary
between gravitationally bound and unbound systems~\citep{clocks}. The
local geometry in the centre of voids is similarly assumed to be given
by a negatively curved FLRW geometry with scale factor $a_{\rmn{v}}$, and
a time parameter $\tv$ which does not in general coincide with $\tw$.

The ensemble of void and wall regions is assumed to evolve by a Buchert
average \citep{Buchert00}, with a volume--averaged scale factor
\beq\ab^3 =\fvi\av^3 + \fwi\aw^3\equiv\ab^3(\fv+\fw),\eeq
where $\fvi\ll1$ is the initial void fraction, and $\fwi=1-\fvi$. The
Buchert equations determine $\ab^3$ in terms of an average volume expansion
on spatial hypersurfaces comoving with the dust. Consequently the scale
factor $\ab$ is a statistical quantity which cannot be directly assigned
any local geometrical meaning. It is therefore impossible to apply solutions
of the Buchert equations to observed quantities to infer cosmological
parameters until one provides:
(i) an operational definition of what is to be understood by ``dust''; and
(ii) an operational means of relating a local average geometry such as
(\ref{eq:wallmetric}) to the volume-average $\ab$.

The operational interpretation of the Buchert formalism is not directly
addressed in Buchert's original work \citep{Buchert00}, although the broad
issues were discussed by \citet{BC1,BC2}. In the absence of a direct
operational interpretation many authors have simply treated $\ab$ as
if it were the geometrical scale factor in an FLRW model, with corresponding
cosmological parameters\footnote{See, e.g.,
\cite{BLA,Rasanen06,Rasanen08,Mattsson09,LABKC,WB10,Mattssons10}.}.
However, in the presence of strong inhomogeneities every
observer cannot be the same average observer, and variance in local geometry
can play an important role in parameter fitting. With this in mind,
\citet{clocks,sol,equiv,obs} adopts a fundamentally different approach to
the operational interpretation of solutions to the Buchert equations.

Firstly, it is assumed that dust particles are
coarse--grained on scales of order $100/h$ Mpc, on which there are no
appreciable average mass flows. Although atomic sized dust is a perfectly
valid description for cosmology at the epoch of last scattering, once stars and
galaxies form the dust in cosmological models is typically coarse--grained
on scales larger than that of galaxies, so that one need only deal with
regions with the same sign of the expansion rate, neglecting the problems
that arise when particle geodesics intersect and pressure becomes significant.
By coarse--graining on scales on which average mass flows are not appreciable,
we can consistently consider cosmic evolution from last scattering until the
present epoch without having to consider energy fluxes which are not
included in the energy--momentum tensor in Buchert's scheme.

Given the coarse scale on which dust ``particles'' are defined, we interpret
Buchert's time parameter, $t$, as a collective degree of freedom of a dust
particle, which does not necessarily coincide with the clock of an ideal
isotropic observer, namely one who detects an isotropic cosmic microwave
background (CMB). In particular, it is assumed that on account of quasilocal
gravitational energy gradients isotropic observers in galaxies and voids
have clock rates that develop a significant variance cumulatively over the
lifetime of the universe. This means hypothetical observers in voids
unbound to any structures will ultimately measure a different mean CMB
temperature from observers in galaxies. However, it does not affect
cosmological observations directly, since on account of structure formation
we and all the objects we observe are necessarily in regions of greater than
critical density, which are contained within walls and keep a cosmic time
in step with ours.

The Buchert time parameter, $t$, is then that carried by the clock of an
isotropic observer in a region within a dust particle
where the local spatial curvature happens to coincide with the
volume--average spatial curvature. In a universe dominated in volume
by voids, this will necessarily be in a void, though not at a void centre.
The proper time of isotropic wall observers, $\tw$, is related to $t$
by the lapse parameter $$\gb=\frac{\rmn{d}t}{\rmn{d}\tw}.$$
Although backreaction, arising from the variance of the relative expansion
rates of the voids and walls with respect to any one set of
clocks, is necessary to obtain apparent cosmic acceleration, for realistic
cosmological parameters \citep{clocks,LNW,obs} the backreaction is relatively
small as a fraction of the total energy density ($< 5\%$). Thus in contrast to
other approaches to Buchert averaging \citep{Buchert08}, apparent acceleration
is not derived from backreaction alone. The crucial feature which leads to
apparent acceleration for realistic cosmological parameters is the difference
in clock rates between wall observers and the volume average, which can
typically grow to the order of 38\% by the present epoch \citep{LNW}.

To complete the operational interpretation of the solutions to the
Buchert equation, the volume average scale factor $\ab$ is adapted
to a spherically symmetric average metric \citep{clocks}
\beq\label{eq:baremetric}
\dd s^2=-\dd t^2+\bar{a}^2(t)\dd\etb^2 + A(\etb,t)\dd\OM^2,
\eeq
in terms of the Buchert time parameter, where the area function $A$ is
defined by an average over the particle horizon volume~\citep{clocks}.
An effective radial null cone average in terms of the parameters of the
wall geometry is then obtained by conformally matching radial null
geodesics of (\ref{eq:wallmetric}) and (\ref{eq:baremetric}) adapted to a
common centre, to relate the conformal time parameters $\etw$ and $\etb$
according to
\beq
\dd\etw= {\fwi^{1/3}\dd\etb\over\gb\fvf^{1/3}}
\label{2eta}\eeq

In place of (\ref{eq:baremetric}), we then extend the metric
(\ref{eq:wallmetric}) beyond finite infinity to obtain a spherically
symmetric average cosmological metric adapted to wall clocks,
\beq\label{eq:dressedmetric}
\dd s^2=-\dd\tw^2+a^2[\dd\etb^2 +\rw^2(\etb,\tw)\dd\OM^2]
\eeq
where $a\equiv\ab/\gb$, and
\beq\rw\equiv\gb\fvf^{1/3}\fwi^{-1/3}\etw(\etb,\tw),\label{rw}\eeq
with $\fv\equiv \fvi\av^3/\ab^3$.
It should be stressed that neither (\ref{eq:baremetric}) nor
(\ref{eq:dressedmetric}) are exact solutions of the Einstein equations,
but are effective average spherically symmetric geometries that
represent a solution of the statistical Buchert average of the Einstein
equations, when adapted to different clocks and rulers.

Given the two metrics (\ref{eq:baremetric}) and (\ref{eq:dressedmetric}),
there are also different sets of cosmological parameters.
The independent Buchert equations \citep{Buchert00},
\bea
&&\OMM+\OMk+\OMQ=1,\label{Beq1}\\
&&\ab^{-6}\pt_t\left(\OMQ\bH^2\ab^6\right)+\ab^{-2}\pt_t\left(\OMk\bH^2\ab^2
\right)=0,\label{Beq2}
\eea
deal with the volume--average or bare parameters, relative to the metric
(\ref{eq:baremetric}). Here
$\OMM=8\pi G\rhb\Z{M0}\ab\Z0^3/[3\bH^2\ab^3]$,
$\OMk=-\kv\fvi^{2/3}\fv^{1/3}/[\ab^2\bH^2]$, and
$\OMQ=-\dot\fv^2/[9\fv(1-\fv)\bH^2]$ are the
bare matter, curvature and kinematic back-reaction parameters respectively,
$\bH\equiv\dot\ab/\ab$ is the volume--average
or bare Hubble parameter, an overdot denotes a derivative w.r.t.\
volume--average time, $t$, and the average curvature is due to the voids only,
which are assumed to have $\kv<0$. The bare Hubble parameter satisfies
the relation
\beq \bH=\fv\Hv+\fw\Hw,\eeq
where $\Hv\equiv\dot a_{\rmn{v}}/\av$ and $\Hw\equiv\dot a_{\rmn{w}}/\aw$ are
the regional average expansion rates of voids and walls, with respect to
volume--average time. It is convenient to define $\hr(t)\equiv\Hw/\Hv<1$.

Relative to the metric (\ref{eq:dressedmetric}), one can also define
dressed parameters, which will take numerical values similar to those of
cosmological parameters in the FLRW models. Such parameters do not satisfy
a relation such as (\ref{Beq1}), however. The most relevant are the
dressed matter density parameter
\beq\OM_{\rmn{m}}=\gb^3\bar{\OM}_{\rmn{m}},\eeq
and the dressed Hubble parameter
\beq
\Hh\equiv{1\over a}{\dd a\over\dd\tw}=\gb\bH-\Dts\gb=\gb\bH-\gb^{-1}\Dtc\gb\,.
\eeq
The present epoch value of the dressed Hubble parameter will coincide with
the standard Hubble constant, $\Hm$, that we infer observationally from
measurements averaged over both voids and walls on scales larger than the
scale of statistical homogeneity.

One feature of the TS model that is particularly relevant for the analysis
of luminosity distance data is that below the scale of statistical homogeneity
we will expect to see significant variance in the Hubble flow. With respect to
any one set of clocks the underdense voids expand faster than the more dense
walls. Since voids dominate the volume of the universe, if we average only
on nearby scales we will typically measure a higher value of the Hubble
constant than the global average, $\Hm$. A measurement confined to our
own local wall, e.g., towards the Virgo cluster, would produce a lower
value\footnote{While this agrees with observation, astronomers typically
talk about ``Virgo--centric infall'' rather than the expansion rate being
less in regions of greater density. It is our view that such language,
which derives from a conceptual framework of Newtonian gravitational forces
in a static space, is inappropriate on scales over which the volume of space
is expanding and therefore not well described by a static Euclidean
geometry. One should not talk about ``infall'' unless the distance between
two objects is actually decreasing.}. As we look out to greater and greater
distances, a typical line of sight will eventually intersect as many walls
and voids as the global average. Suppose that we perform a spherically
symmetric average over the sky to try to determine the Hubble constant using
only nearby measurements within some given finite maximum radius, which is
then varied. The Hubble ``constant'' inferred in this manner should peak at
the dominant void scale $30/h$ Mpc, i.e. at $z\goesas0.01$, with a maximum
value up to 17\% greater than the global average and then steadily
decrease to near the global average value when the scale of statistical
homogeneity is reached. The latter scale, which coincides with our scale for
the coarse--graining of dust, is about $100/h$ Mpc. These expectations are
consistent with the recent data analysis of \citet{LS}, and future
observations would have the potential to either rule out or strongly
constrain the TS scenario.

Since the scale of statistical homogeneity represents a redshift of order
$z\goesas0.033$, and since many supernovae datasets contain a large number
of supernovae below this scale, there are many potential systematic
issues to consider. Such issues were not considered by \citet{KFL},
and will represent an important ingredient in our reanalysis.

Provided we are sampling distance scales {\em beyond} the scale of statistical
homogeneity, then the luminosity distance for wall observers is
given by~\citep{clocks}
\beq\label{eq:tslumdist1}
\dL=a\Z0(1+z)\rw,
\eeq
where $\rw$ may be determined from (\ref{2eta}) and (\ref{rw}).

The general solution for the two--scale average of the Buchert equations
was given by \citet{sol}, and is expressed in terms of transcendental
equations. It has four free parameters, two of which may be expressed
as the initial void fraction, $\fvi$, and the initial velocity dispersion
$\hr(\ti)$. For initial conditions at last scattering consistent with
the CMB we take initial values $10^{-5} < \fvi < 10^{-2}$, $1-\hr(\ti)=
10^{-5}$. However, it turns out that the general solution is relatively
insensitive to these values, as there is an attractor solution with $\hr(t)
=\frac23$ exactly, which the general solutions approach to within 1\% by
redshifts $z\goesas37$ \citep{sol}. The solutions then depend effectively
on two free parameters, which can be taken to be the dressed Hubble constant
and dressed matter density parameter at the present epoch.

For the tracker solution the dressed matter density parameter and void
fraction at the present epoch are related by
\beq\label{eq:fv0Om}
\OM_{\rmn{m}0} =\frac{1}{2}(1-\fvn)(2+\fvn).
\eeq
Furthermore, for the tracker solution, (\ref{2eta}), (\ref{rw}) and
(\ref{eq:tslumdist1}) simplify to give \citep{obs}
\bea \label{eq:tslumdist}
\dL&=&(1+z)^2t^{2/3}\int_t^{t\X0}
{2\,\dd\tb\over(2+\fv(\tb))(\tb)^{2/3}}\nonumber\\
&=&(1+z)^2{t^{2/3}(\FF(t\Z0)-\FF(t))}
\eea
where
\bea
\FF(t)&\equiv&2t^{1/3}+{b^{1/3}\over6}\ln\left((t^{1/3}+b^{1/3})^2\over
t^{2/3}-b^{1/3}t^{1/3}+b^{2/3}\right)\nonumber\\
&&\qquad+{b^{1/3}\over\sqrt{3}}\tan^{-1}\left(2t^{1/3}-b^{1/3}\over
\sqrt{3}\,b^{1/3}\right),\label{FF}\eea
and $b\equiv2(1-\fvn)(2+\fvn)/(9\fvn\Hb)$. For
the tracker solution wall time is related to volume--average time by
\beq
\tw=\frac23t+{4\OmMn\over27\fvn\Hb}\ln\left(1+{9\fvn\Hb t
\over4\OmMn}\right)\,,\label{tsol}
\eeq
and the bare Hubble constant to the dressed Hubble constant by
\beq
\Hb={2(2+\fvn)\Hm\over4\fvn^2+\fvn+4}\,.
\eeq
In the data analysis that follows, we used both the exact solution
with $\fvi=10^{-4}$ and $\hr(\ti)=0.99999$ at $z=1100$, and the tracker
solution, and found that they gave essentially identical results to the
accuracy quoted.

\section{Supernova Ia data reduction methods} \label{sect:fitters}
In this paper we consider two \sn\ light curve fitters. Current fitting
methods descend from those used in the original discoveries of cosmic
acceleration in 1998/1999: The Multicolor Light Curve Shape fitter
MLCS2k2~\citep*{JRK} is the most recent incarnation of the LCS fitter used
by the High-Z Supernova Team~\citep{Riess98}, and the SALT/SALT II (Spectral
Adaptive Light curve Template) fitters~\citep{Guy05,Guy07} improve the
magnitude calibration based on the width-luminosity relation used by the
Supernova Cosmology Project~\citep{P98}.

Each method results in a distance modulus for each supernova. However,
distance moduli computed for the same objects by the two fitting methods are
not necessarily equal, due in large part to the different treatment of
systematic uncertainties -- in particular, colour variation due to dust
extinction. It is
very important when estimating cosmological parameters from the data from
various \sn\ observation programmes that the data reduction process is
consistent across the whole dataset.

When one is investigating an alternative cosmological model with \sne, it
is also vital to recognise that model dependence is introduced into the
cosmological parameter estimation at different points in the data reduction
process. We now compare the MLCS2k2 and SALT/SALT II algorithms in preparation
for using their output distance moduli to test the timescape model. More
thorough comparisons of these methods can be found in~\citet{Hicken09} and
in~\citet{Kessler09}.

\subsection[]{MLCS2k2}
For each supernova, MLCS2k2 returns a distance modulus $\mu=5\log_{10}(d_L/
10~\rmn{pc})$ and its uncertainty via the MLCS2k2 model magnitude~\citep{JRK}
\bea\label{mlcs}
m_{\rmn{model}}^{e,f} = M^{e,f'} + \mu + p^{e,f'}\Delta + q^{e,f'}\Delta^2
\nonumber\\
+ X_{\rmn{host}}^{e,f'} + X_{\rmn{MW}}^{e,f} + K^e_{f,f'}.
\eea
A model magnitude is fitted at each time point (indexed by $e$, and defined
relative to the time of maximum brightness), and for each observer-frame
filter index $f$. The model is defined in {\em UBVRI} rest frame filters $f'$,
with the light curve shape parameter $\Delta$ representing the relation
between luminosity and duration. Extinction due to dust is divided into host
galaxy $X_{\rmn{host}}$ and Milky Way $X_{\rmn{MW}}$ components, and
$K_{f,f'}$ is the K-correction between rest-frame and observer-frame filters.
Obtaining a set of distance moduli takes two steps. One first computes the
model vectors $M^{e,f'}$, $p^{e,f'}$, and $q^{e,f'}$ by minimising the distance
modulus residuals of a training set of nearby \sn, which lie within the
range in which the Hubble line is linear, yet are sufficiently distant for
their peculiar velocities to be negligible compared with their Hubble-flow
velocity. Secondly, one fits for the distance moduli along with the remaining
parameters, on the assumption that \sne\ at higher redshifts are identical to
those nearby.

The extinction terms are set up independently of these fits. There is some
colour dependence on the brightness incorporated in $\Delta$, but there is
also a significant colour dependence on extinction by dust in the host galaxy
and in the Milky Way. Following~\citet*{CCM}, MLCS2k2 characterizes the
extinction by first defining $\zeta^{e,f'}\equiv X^{e,f'}/A^{e,f'}_0$, where
$A^{e,f'}_0$ is the extinction in passband $f'$ at maximum light in $B$. Hence,
$\zeta^{e,f'}(t=0)$ is defined to be 1, and $\zeta^{e,f'}$ encapsulates the
time dependence of the exinction. Then one fits for the coefficients
$\alpha^{f'}$ and $\beta^{f'}$, defined by
\beq
X^{e,f'}=\zeta^{e,f'}\Big(\alpha^{f'} + \frac{\beta^{f'}}{\RV}\Big)A^0_V,
\eeq
at maximum light, where $\RV$ is the ratio of the extinction in the $V$-band
to the colour excess $E(B-V)$. A ``Milky Way-like'' reddening law, based on
an average over a number of lines of sight, has $\RV=3.1$. This has
conventionally been assumed to apply also in SN Ia host galaxies. However,
\citet{Hicken09} and~\citet{Kessler09} find that this value for $\RV$
\emph{overestimates} host galaxy extinction. The question of how to
parameterise the extinction is complicated by a degeneracy with Hubble bubble
assumptions~\citep{Conley07}. Different groups use different values.

There are four model parameters for each SN Ia: the distance modulus $\mu$;
the time of peak luminosity in the rest-frame $B$-band; the
shape-luminosity parameter $\Delta$; and host galaxy extinction $A_V$. The
estimates and uncertainties of each parameter value are determined by
marginalizing over the three other parameter probability distribution
functions and taking the mean and rms of the resulting probability
distribution for the parameter of interest.

With a distance modulus and its uncertainty for each supernova, cosmological
parameter estimates are obtained by minimizing the $\chi^2$ statistic:
\beq\label{eq:chi2}
\chi^2=\sum_i\frac{[\mu_i - \mu(z_i,\Hm,\OmMn)]^2}{\si_{\mu_i}^2}
\eeq
where $\mu(z_i,\Hm,\OmMn)$ is the theoretical distance modulus calculated for
the redshift of the $i$-th \sn, based on a set of cosmological parameters
$\Hm$ and \Omm\ for a spatially flat universe.

\subsection[]{SALT/SALT II}
Whereas the MLCS calibration uses a nearby training set of SNe assuming
a close to linear Hubble law,
SALT~\citep{Guy05} uses the whole dataset to calibrate empirical light
curve parameters. \sne\ from beyond the range in which the Hubble law is
linear are used, so a cosmological model must be assumed in this method.
Typically the \LCDM\ model is assumed. To deal with a determination of
empirical parameters for objects at unknown distances, the absolute magnitude
$M$ and $\Hm$ are combined as
\beq\scrM=M-5\log_{10}h+5\log_{10}c+10.\label{scrM}\eeq
The distance modulus is then modeled as
\beq\label{eq:salt}
\mu_i = m_{B_i}^{\rmn{max}} - \scrM + \alpha(s_i - 1) - \beta c_i.
\eeq
The initial light curve standardization results in best fit values for the
time of maximum $B$-band light, $t_0$, the rest-frame peak magnitude in the
$B$-band, $m_B^{\rmn{max}}$, a stretch factor $s$, and a colour parameter $c$,
in which are combined the intrinsic supernova colour and reddening due to
dust in its host galaxy.

SALT II~\citep{Guy07} builds on SALT by including spectroscopic
information to improve the wavelength resolution of the spectral templates. We
use the relationship between the SALT stretch parameter $s$ and the SALT II
parameter $x_1$ given in \citet{Guy07} in order to compute cosmological
parameters for SALT II \sne\ with the same program as we use for the SALT \sne.

In MLCS2k2, the cosmological parameter estimation step
is ``decoupled'' from the distance modulus determination. In SALT, after
obtaining the parameters $s$ and $c$ for each \sn\ from the light curve fits,
a \emph{magnitude} for each supernova is fitted via the equivalent expression
to eq.~(\ref{eq:chi2}), and the cosmological parameters are obtained as part
of the same minimization, viz.
\beq\label{eq:saltchi2}
\chi^2=\sum_i\frac{[m_{Bi} -
m(z_i;\alpha,\beta,\scrM,\OmMn)]^2}{\si_{m_i}^2}.
\eeq
This process results in global estimates of $\alpha$, $\beta$, and $\scrM$, and
a corresponding \Omm. An additional ``intrinsic'' dispersion is introduced to
the \sn\ absolute magnitudes such that one obtains a reduced $\chi^2$ of
1 for the best fit set of parameters \citep{Guy07}. Consequently the published
tables of SN Ia distance moduli obtained with the SALT/SALT II fitters retain a
degree of model dependence. In the case of the Union and Constitution datasets,
the quoted distance moduli were
computed for a spatially flat (FLRW) universe with a constant $w$.

There are many subtleties in the individual implementations of the \sne\
reduction. The cosmological parameters computed for the Union dataset of
\citet{Kowalski08} are obtained from the $\chi^2$ equation~(\ref{eq:saltchi2})
above, but with additional nuisance parameters encoding the propagating
systematic uncertainties (see their equation (5)).

\subsubsection[]{Combining datasets}

Since SALT uses the whole dataset, the global parameter minimization needs
to be rerun when new data is added. When they augment the Union set with the
nearby CfA3 \sne\ to produce the Constitution dataset, \citet{Hicken09}
first take the SALT parameters of the Union dataset, $m_B^{\rmn{max}}$, $s$,
and $c$, ``out of the box'' and calculate a best-fit cosmology incorporating
a BAO prior in the cosmological fit: $\chi^2=\chi^2_{\mu}+\chi^2_{\rmn{BAO}}$.
The reason for this was presumably that the BAO prior constrains the range of
\Omm\ better than the \sne\ alone, and thus provided a more stringent
assurance that the new cosmological parameter estimation was in line
with the old one. Once the Union results were reproduced with sufficient
accuracy, the light curves of nearby Union \sne\ were run once again
through SALT to ensure that the new $m_B^{\rmn{max}}$, $s$, and $c$ values
agreed with the Union ones, and then the whole CfA3 sample was run through
SALT, so that it could be combined with the Union set without introducing any
significant offset. The uncertainty in the distance modulus $\si_{\mu}$ was
calculated by \citet{Hicken09} in a way that differs from \citet{Kowalski08}.
Essentially, the $\si_{\mu}$ calculated by \citet{Hicken09} contains
adjustments that ensure reproduction of the same uncertainty in $w$ as in
\citet{Kowalski08}.

If one wishes to test a non-standard cosmological model using SALT \sne, as in
the present study, the minimization process in eq.~(\ref{eq:chi2}) should at
the very least be recalculated with the appropriate luminosity distance
formula to determine the empirical light curve parameters. However, for
consistency, the assumptions underlying the determination of the uncertainties
in the $\chi^2$ minimization procedure should ideally also be carefully
rethought. The analysis must be consistent across both standard and
non-standard cosmological models in order to produce a meaningful Bayes
factor. From the standpoint of the non-standard model, when combining datasets
as above one can no longer simply manipulate the uncertainties in such a way
that published constraints for the parameter $w$ of the standard cosmology
are more or less reproduced.

\subsection[]{Systematic uncertainties}

It has been noted in several studies, (e.g., \citet{Hicken09,Kessler09,%
Komatsu2010}), that there are possible discrepancies between SALT- and
MLCS-reduced datasets, and also between different implementations of the
same fitters. This is a significant issue for our investigation.

\begin{figure}
\begin{center}
\caption{Differences in published distance moduli between the 140
\sne\ common to the Riess07 gold sample and the Constitution sample, as
a function of redshift.
\label{fig:140}}
\includegraphics[scale=0.3,angle=270]{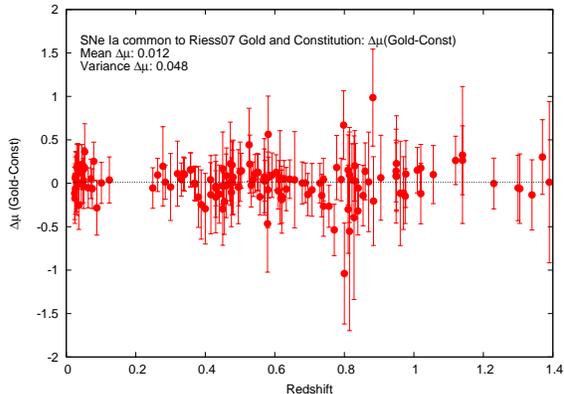}
\end{center}
\end{figure}

One direct way of quantifying the differences is to compare the published
distance moduli for the 140 \sne\ common to the Riess07, Union and
Constitution datasets. In Fig.~\ref{fig:140} we plot the differences between
the published Riess07 and Constitution distance moduli for the 140 data points
they have in common. This shows that
$|\mu_{\rmn{Gold}}-\mu_{\rmn{Const}}|\lesssim 1$~mag.
Individually these differences are quite significant in some cases.
However, the differences show no obvious redshift-dependent trend so
they should not bias the relative estimates of cosmological parameters.

\begin{figure}
\begin{center}
\caption{Differences in distance moduli between values published in
the Constitution sample, and the corresponding values determined with
\cosfit\ (\LCDM) using only the subsample of 140 \sne\ plotted in
Fig.~\ref{fig:140}, as a function of redshift.
$\Hm=65$ \kmsmpc.\label{fig:140sc}}
\includegraphics[scale=0.3,angle=270]{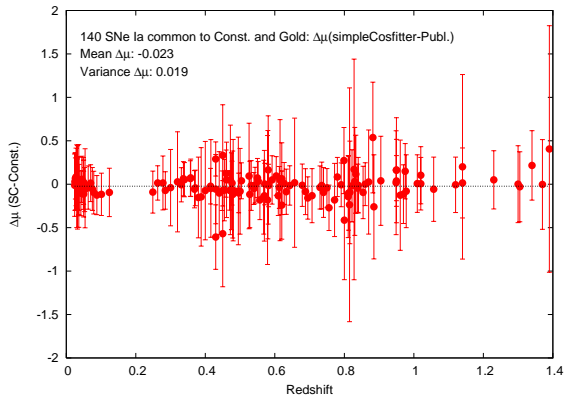}
\end{center}
\end{figure}

\begin{figure}
\begin{center}
\caption{Differences in distance moduli between published values in the
Constitution sample, and values determined with \cosfit\ (\LCDM)
using the full Constitution sample, as a function of redshift.
$\Hm=65$ \kmsmpc.\label{fig:chk}}
\includegraphics[scale=0.3,angle=270]{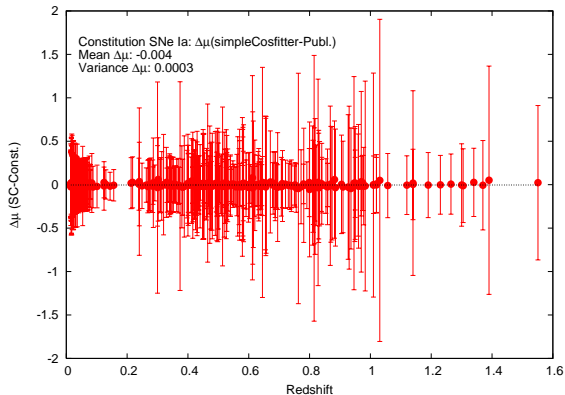}
\end{center}
\end{figure}

One should be careful not to draw the conclusion that the differences
seen in Fig.~\ref{fig:140} are only to be found in comparing MLCS (used
in the Riess07 gold sample) with SALT (used in the Constitution sample). As
discussed above, when using the SALT fitter a global refit of all the empirical
light curve and cosmological parameters is required when extra data is
included in a sample. To illustrate the effect of this, we have taken the same
140 \sne\ common to the Riess07, Union and Constitution datasets and have
computed a SALT fit to this subsample directly, assuming a spatially flat
\LCDM\ model, using Conley's publicly available \cosfit\
code\footnote{\tt http://qold.astro.utoronto.ca/conley/simple$\_$cosfitter/}.
In Fig.~\ref{fig:140sc} we display the differences between these distance
moduli and the corresponding distance moduli of the same \sne\ events as
published in the Constitution compilation. We see that the resulting scatter
is half of that in Fig.~\ref{fig:140}, even though the SALT method
has been applied in each case. This is due to the variation in the values
of the parameters $\alpha$, $\beta$ and $M_B$ determined from only 140 \sne\
in the subsample, as compare to those determined from all 397 \sne\ in the
full sample.\footnote{Since the \cosfit\ implementation of SALT is not
perfectly identical to that used by \citet{Hicken09}, as a check we reran
\cosfit\ on the full Constitution sample. The difference between the
published fits and those computed with \cosfit\ are indeed insignificant,
as is shown in Fig.~\ref{fig:chk}.}

Much more detailed studies of these sorts of issues have already been
conducted by various researchers, revealing a complex picture.

In their comparison of the different fitters, \citet{Hicken09} found that
SALT, SALT II, and MLCS2k2 produce light curve shape and color/reddening
parameters that agree well with each other and that it is in determining
the distance modulus that systematic offsets are introduced. For example,
SALT produces more scatter at high redshifts than MLCS2k2, and the nearby
MLCS17 distances are larger than in SALT. Such discrepancies will clearly
affect the cosmological parameter fits. \citet{Hicken09} note that although
there exist some trends in the $\mu$(MLCS17)$-\mu$(SALT) differences versus
shape parameter $\Delta$ and color parameter $\beta$, there are no trends
versus redshift, which would indicate the influence of hidden systematics
and affect the cosmological parameter fits~\citep{wv07}. These are grounds
for hope that retraining with larger datasets, combined with a better
treatment of systematic uncertainties, will reconcile the differences. In
the meantime, however, they found that for the best cut samples, SALT and
SALT II estimates of $w$ differ by 0.05--0.09 from the MLCS estimates.

After accounting for an anomaly in the rest-frame $U$-band that affects the
nearby \sne\ with the SALT II fitter and all except the nearby sample with the
MLCS2k2 fitter, \citet{Kessler09} found a discrepancy of 0.31 in the value
of $w$ estimated from one of their MLCS2k2 sample combinations as compared
to that of the corresponding sample from the MLCS2k2 ESSENCE
collaboration\citep{wv07}. They show that this is due to different
parameterisations of the dust extinction term in different implementations of
MLCS2k2.

When constraining dark energy parameters derived from
WMAP7+BAO+SN,~\citet{Komatsu2010} found that the parameters of the minimal
6-parameter \LCDM\ model changed depending on whether the \sn\ compilation used
SALT II or MLCS2k2 based on the SN sample of \citet{Kessler09}. They noted
that it is not presently obvious how to properly incorporate systematic
uncertainties into the likelihood analysis and thereby reconcile different
methods and datasets. \citet{Komatsu2010} use the Constitution sample when
quoting canonical cosmological parameter values, because it is an extension of
the Union sample which they used for the 5-year WMAP analysis.

\subsubsection[]{The Hubble bubble\label{sect:bubble}}
In addition to the general question of the effect of unknown systematics
on cosmological parameters, there is one particular systematic which is
of interest for the TS model, namely the possible existence of a
Hubble bubble. As discussed in Sec.~\ref{sect:ts} it is a feature
of the TS model that we will observe an apparent increase in the value of
the Hubble constant on scales less than the scale of statistical
homogeneity at $z\sim0.033$. A spherically symmetric average of the
Hubble rate over increasing local volumes will give a peak variance of
order 17\% above the mean on scales $z\goesas0.01$, which then steadily
decreases until the scale of statistical homogeneity is reached.

Usually, very low redshift objects are left out of \sne\ samples because
their peculiar velocities are a considerable fraction of their Hubble flow
velocities, but depending on the sample events with redshifts $z\gsim0.01$
have been included. Evidence of a Hubble bubble was found by
\citet{Zehavi98}, and confirmed by \citet{JRK}, using a MLCS2k2 sample with
$\RV=3.1$. Modeling the expansion law in terms of a single inner region void
expanding faster than the outer FLRW region, they found a drop in the Hubble
constant of $\de\Z H=(H\ns{inner}-H\ns{outer})/H\ns{outer}=0.065$ at a
redshift $z=0.024$. However, there exists a degeneracy between the existence
of such a small scale Hubble bubble and the treatment of
reddening/extinction~\citep{Conley07}.

In the SALT-reduced samples, a Hubble bubble is found if $\beta=4.1$ is
assumed -- which is believed to roughly correspond to the CCM89 value for
reddening in the Milky Way -- but disappears if $\beta<4.1$ ~\citep{Conley07}.
With their MLCS31 sample (366 \sne) \citet{Hicken09} find $5.56\si$ evidence
for a void at $z=0.028$ with reduced amplitude $\de\Z H=0.029$. In their
MLCS17 sample (372 \sne) by contrast, with $\RV=1.7$, they find $2.75\si$
evidence for a negative void at $z=0.034$ with $\de\Z H=-0.020$.

It is clear that unknown systematic uncertainties in reddening and
extinction of supernovae in their host galaxies will lead to different
results regarding local inhomogeneities. Different groups have made
different choices about the minimum cutoff in light of these uncertainties.
\citet{Riess06} took a minimum redshift $z=0.024$, whereas \citet{Hicken09}
included data down to $z=0.01$. By contrast, \citet{Kessler09}, for their full
MLCS2k2 Nearby+SDSS+SNLS+ESSENCE+HST sample, took a minimum
redshift of $z=0.0218$, based on the midpoint of a $\pm0.06$ variation in $w$
with minimum redshift cuts between 0.01 and 0.03.

The statistical nature of the apparent Hubble expansion law expected
in the TS model will differ from the empirical models used in the above
analyses of the Hubble bubble, as we are not dealing with a uniformly
expanding void inside a background region. However, in general an increased
minimum redshift cut\footnote{The effects of minimum redshift cuts have
also recently been discussed for the Hubble bubble generated by single--void
Lema\^{\i}tre--Tolman--Bondi models \citep{Sinclair10}. This will also have
different characteristics to the apparent Hubble bubble in
the TS model.} should be made in the TS model, a point to which
we will return in Sec.~\ref{sect:shs}.

\subsection[]{The value of $H_0$ and cosmological fits} \label{sect:H0}
The overall normalisation of the luminosity distance depends on the value of
the Hubble constant, $\Hm$. However, one cannot extract information about the
value of the Hubble constant independently of a knowledge of the intrinsic
luminosity of standard candles, since uncertainties in the parameters $M$
and $\Hm$ are degenerate with one another in the distance modulus. The SALT
fitter provides a global estimate of $\scrM$ in which any uncertainties
are combined according to (\ref{scrM}), and so say nothing about the value
of $\Hm$ directly. The value of $\Hm$ must be determined by an independent
calibration.

In the MLCS method, the overall distance scale similarly relies on the
calibration of the magnitudes of nearby \sne, usually to
the Cepheid distance scale~\citep{Freedman00,Sandage06,shoes}.

It is impossible therefore to infer the value of the Hubble constant
by a fit to \sne\ data alone. However, for the MLCS method, in which
the uncertainties in the intrinsic magnitudes have hopefully already
been accounted for, the fit of luminosity distances to a particular
cosmological model can nonetheless provide an estimate of the variance
in $\Hm$ values that are admissible for that model, given a particular
\sne\ dataset. Since independent cosmological tests, such as the
determination of the angular diameter distance of the sound horizon,
or of the comoving baryon acoustic oscillation scale, lead to different
constraints on $\Hm$, in order to compare the potential agreement of
different tests \citet{LNW} plotted confidence levels for the fit to
the Riess07 gold data in their Fig.~2, using the normalization assumed
in the data\footnote{\citet{Riess06} did not state what value of $\Hm$
was assumed in their dataset, but stated that a systematic subtraction of
0.32 mag from their distance moduli would match the Cepheid calibration
of \citet{Riess05}. The question of the Cepheid calibration is a matter
of debate \citep{Sandage06,shoes}. Since a fit of the spatially flat \LCDM\
model to the unmodified Riess07 \sne\ distance moduli gives a value
$\Hm=62.6\pm1.4\kmsMpc$ consistent with the value
determined by \citet{Sandage06}, \citet{LNW} chose to use the unmodified
distance moduli of \citet{Riess06}. The recent Cepheid calibration
of \citet{shoes} is in disagreement with the \citet{Sandage06} calibration.}.
One should bear in mind that these confidence limits can be translated
up or down the $\Hm$-axis, depending upon what overall normalization is assumed
for the Hubble constant.

\begin{figure}
\begin{center}
\caption{Parameter values in the ($\OmMn$,$\Hm$) plane which fit the angular
scale of the sound horizon $\de=0.01$ rad deduced for WMAP
\citep{wmap1,wmap3}, to within 2\%, 4\% and 6\% (contours running top--left
to bottom--right); as compared to parameters which fit the effective comoving
scale of $104h^{-1}$Mpc for the baryon acoustic oscillation (BAO) observed in
galaxy clustering statistics \citep{Cole,Eisenstein}, to within 2\%, 4\% and
6\% (contours running bottom--left to middle--right)
which fit the angular diameter dis
of Gold07 (Table \ref{tab:cutmlcs17}), SDSS-II (Table \ref{tab:hubblebubble}),
MLCS17 and MLCS31 (Table \ref{tab:mlcscommon}). In each case an overall
normalization of the Hubble constant from the published dataset is assumed.
\label{fig:cmbbao}}
\includegraphics[scale=0.48]{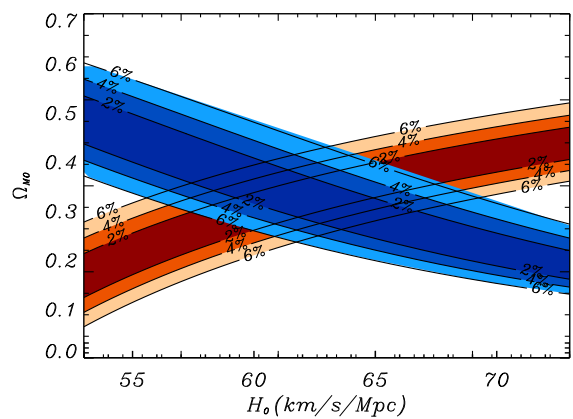}
\end{center}
\end{figure}

Constraints on the Hubble constant from WMAP and baryon acoustic oscillations
in the TS model were given in Fig.~2 of \citet{LNW} and are reproduced in
Fig.~\ref{fig:cmbbao}. The constraints from WMAP are estimated by fitting
the angular scale of the sound horizon to within 2, 4 or 6\%. The BAO
constraints are similarly estimated by assuming that the dressed comoving
scale of the sound horizon matches the corresponding scale of $104\,h^{-1}$
Mpc for the \LCDM\ model to within 2, 4 or 6\%. Assuming that these estimates
roughly correct\footnote{A direct comparison with the data requires that
we compute the expected angular anisotropy power spectrum for the TS model
in the case of the CMB, and also that all model dependent assumptions in
the data reduction of galaxy clustering data are carefully re-examined in
the case of the BAO analysis.} then Fig.~\ref{fig:cmbbao} shows that values of
$57\lsim\Hm\lsim68\kmsMpc$ would be admissible in the TS scenario, but values
as large as the recent $\Hm=74.2\pm3.6\kmsMpc$ determined by the SH0ES survey
\citep{shoes} would represent a severe challenge to the model.

The determination of the value of the Hubble constant is a complex
problem from the viewpoint of the TS model, given that many of the crucial
measurements are made below the scale of statistical homogeneity, over
scales on which the local Hubble flow should exhibit quite large variability.
Indeed, the statistical properties of the observed fractional variability
of the Hubble flow \citep{LS} do seem broadly consistent with the
expectations of the TS scenario. For concordance of the geometrical
tests of average cosmological parameters, the real challenge is the
baseline value of $\Hm$, however.

Ideally the global average Hubble constant should be determined only
on scales significantly larger than the scale of statistical homogeneity,
$z>0.033$, by methods which do not depend on calibrations below that scale.
The method of determining $\Hm$ via the time delays of multiply-imaged
quasars in strongly gravitationally-lensed systems fulfils this criteria.
This method has given a large variety of estimates for $\Hm$ \citep{Oguri07}.
A recent new estimate of $\Hm$ from accurate time delay measurements with
six years of data from the quadruply imaged quasar HE 0435-1223 gives
$\Hm=62^{+6}_{-4}\kmsMpc$ \citep{Courbin10}. In considering such estimates
from the perspective of the TS model, one must be careful to examine any
assumptions which might assume the standard cosmology.

The analysis of the Sunyaev--Zel'dovich effect and X--ray data of galaxy
clusters provides another method of constraining $\Hm$ independently of
calibration to standard candles in the extragalactic distance ladder
\cite{Reese}. If the standard \LCDM\ angular diameter distance is replaced
by that of the TS model then this method could be easily adapted to give
further constraints in the $(\Hm,\OmMn)$ parameter space. This is beyond
the scope of the present paper, and will be left to future work.

Since we are not interested in the comparing other cosmological tests
in this paper, we will not consider the question of the value of $\Hm$
further. Rather we will concentrate on the comparison of the expansion
history for the TS and spatially flat \LCDM\ models as determined by the
luminosity distances of various \sne\ datasets.

\section{Supernova analysis of the timescape model} \label{sect:results}

In this section we will test the TS model against all available datasets.
In view of the systematic uncertainties it is important
that we consider the effects of using the different fitters, as well as
additional possible systematic effects specific to the TS model. In
$\S~$\ref{sect:fits}, we discuss TS fits to the published datasets ``out of the
box''. In $\S~$\ref{sect:recal}, we determine the extent to which substitution
of the TS luminosity distance calculation for the \LCDM\ one in the \cosfit\
code affects parameter estimation. We then investigate the effects of making
sample cuts according to the redshift corresponding to the scale of statistical
homogeneity in $\S~$\ref{sect:shs}. Finally, in $\S~$\ref{sect:mlcssys} we
discuss some systematic issues relevant to MLCS2k2.

\subsection[]{``Out of the box'' data} \label{sect:fits}
\citet{KFL} simply took the published values of the Union and Constitution
datasets, produced data fits and concluded that the TS model was a relatively
poor fit, with the implication that the TS failed when presented with the
newer larger datasets. However, as we discussed above, given that the Union
and Constitution datasets are fit by SALT, which implicitly assumes a
homogeneous isotropic cosmology to all distances in its data reduction,
serious concerns arise in using the SALT method. In fact, it was for this
reason that the MLCS--reduced data of Riess07 was used by \citet{LNW} in
preference to SALT--reduced SNLS data.

For the purpose of subsequent comparison, we will first collate TS
cosmological parameter fits for all available datasets in
Table~\ref{tab:outofbox} using the data ``out of the box''. The three
datasets investigated in this manner by \citet{KFL} were the Riess07 gold
data \citep{Riess06}, and the SALT-fitted Union \citep{Kowalski08} and
Constitution \citep{Hicken09} datasets. To these we add the equivalent
parameter fits for the SALT2, MLCS17 and MLCS31 datasets of \citet{Hicken09},
the 288-\sn\ Nearby+SDSS+ESSENCE+SNLS+HST sample of \citet{Kessler09},
and the 557-\sn\ Union2 sample of \citet{Amanullah10}. The MLCS17 and
MLCS31 datasets share many \sne\ in common with the Constitution set, but
were fitted by MLCS2k2 with values for the extinction parameter $\RV$ of 1.7
and 3.1 respectively. The value $\RV=1.7$ was found by \citet*{Hicken09} to
give less scatter in the Hubble residuals for a sample of nearby \sne, in
keeping with previous studies which found the colour parameter $\beta$ in
eq.~(\ref{eq:salt}) to be significantly lower than would be expected if the
host galaxy reddening law conforms to a Milky Way reddening law ($\RV=3.1$).
The SDSS-II data \citep{Kessler09} was fitted by MLCS2k2
with $\RV=2.18$. The Union2 dataset is fit with SALT II.
\begin{table}
\begin{center}
\begin{minipage}[t]{\linewidth}
\caption{Expectation values for the parameters for the timescape model from
SNe Ia data, using published \sne\ data as selected and reduced by the
respective \sne\ collaborations. The bestfit value of \Omm\
is quoted in brackets in addition to the expectation value. The MLCS17,
MLCS31, and SALT2 datasets were published along with the Constitution dataset
in \protect\citet{Hicken09}. The SDSS-II dataset is the full 288-object dataset
described in \protect\citet{Kessler09} and available from
\texttt{http://das.sdss.org/va/SNcosmology/sncosm09$\_$fits.tar.gz}.
The sample size, $N$, and minimum $\chi^2$ are also tabulated. S=SALT;
S2=SALT2; M=MLCS2k2.
\label{tab:outofbox}}
\begin{tabular}{lccccl}
\hline\hline
Dataset &$N$& $\chi^2$ & \Omm & $f_{v0}$ \\
\hline
Riess07 (M)&182& 162.7 & $0.29^{+0.14}_{-0.13}(0.33)$ &
$0.79^{+0.11}_{-0.12}$ \\ [1ex]
Union (S)&307& 319.6 & $0.12^{+0.14}_{-0.12}(0.09)$ &
$0.91^{+0.09}_{-0.08}$ \\
[1ex]
Const.~(S)&397& 470.8 & $0.10^{+0.08}_{-0.09}(0.01)$ &
$0.93^{+0.06}_{-0.06}$ \\ [1ex]
MLCS17 (M)\footnotemark[1]&372& 403.1 & $0.18^{+0.12}_{-0.15}(0.20)$ &
$0.87^{+0.09}_{-0.10}$ \\ [1ex]
MLCS31 (M)\footnotemark[1]&366& 432.8 & $0.07^{+0.04}_{-0.06}(0.01)$ &
$0.95^{+0.02}_{-0.04}$ \\ [1ex]
SALT2 (S2)\footnotemark[1]&352& 346.8 & $0.11^{+0.11}_{-0.10}(0.04)$ &
$0.92^{+0.08}_{-0.07}$ \\ [1ex]
SDSS-II (M)\footnotemark[2]&288& 240.8 & $0.38^{+0.11}_{-0.09}(0.40)$ &
$0.72^{+0.05}_{-0.05}$ \\ [1ex]
Union2 (S2)\footnotemark[3]&557& 550.9 & $0.08^{+0.05}_{-0.07}(0.01)$ &
$0.95^{+0.03}_{-0.04}$ \\ [1ex]
\hline
\end{tabular}
\end{minipage}\hfill
\begin{minipage}[t]{\linewidth}
$^1$\citet{Hicken09}
$^2$\citet{Kessler09}
$^3$\citet{Amanullah10}
\end{minipage}
\end{center}
\end{table}

We have compiled Table~\ref{tab:outofbox} by our own analysis of the data,
with the prior\footnote{\citet{KFL} state that this prior corresponds
to taking a prior $10^{-5}<\fvi<10^{-2}$ on the void fraction at last
scattering. However, while the value of \Omm\ is closely related to $\fvn$,
it is essentially independent of $\fvi$ on account of the existence of the
tracker solution. \citet{LNW} used the value $\fvi=10^{-4}$ with the
exact solution for {\em all} values of \Omm; the value of \Omm\ is
essentially insensitive to the value $\fvi$.}
$0.01\le\OmMn<0.95$ used by \citet{KFL}.
For comparison, Table~\ref{tab:lcdmbestfit} shows \LCDM\
parameter values that were published with the respective datasets.
\begin{table}
\begin{center}
\begin{minipage}[t]{\linewidth}
\caption{Published values for \Omm\ for the \LCDM\ model from SNe Ia data, for
comparison with Table~\ref{tab:outofbox}. The Constitution
value of \Omm\ includes a BAO prior. The SDSS-II value includes BAO and WMAP5
priors. Distance normalisation is arbitrary. \label{tab:lcdmbestfit}}
\begin{tabular}{lcccl}
\hline\hline
Dataset & $N$ & $\chi^2$ & \Omm\\
\hline
Union & 307 & 310.8 & $0.29^{+0.05}_{-0.04}$\\ [1ex]
Constitution & 397 & --- & $0.28^{+0.04}_{-0.02}$\\ [1ex]
SDSS-II\footnotemark[1] & 288 & 237.9 & $0.31^{+0.02}_{-0.02}$\\ [1ex]
Union2\footnotemark[2] & 557 & --- & $0.274^{+0.040}_{-0.037}$\\[1ex]
\hline
\end{tabular}
\end{minipage}\hfill
\begin{minipage}[t]{\linewidth}
$^1$These values come from table 13 in \citet{Kessler09}.
$^2$Statistical and systematic uncertainties combined.
\end{minipage}
\end{center}
\end{table}

The parameter values quoted by \citet{LNW} and \citet{KFL} were those
corresponding to the peak in the probability distribution, at which the
$\chi^2$ statistic is minimised. However, as Fig.~1 of \citet{KFL}
illustrates, for the published Union and Constitution datasets the bestfit
value is driven to the edge of the parameter space at
unrealistically small values of \Omm, an issue we will discuss further in
Sec.~\ref{sect:shs}. Given probability distributions such as these that are
significantly skewed relative to a Gaussian distribution, a more typical
estimate of any parameter $\th$ is given by the expectation value
$\langle\th\rangle=\int_{-\infty}^{+\infty}\th\,w(\th)\,\dd\th$, where
$w(\th)$ is the normalised weight. While the bestfit value gives the most
probable individual parameters, the expectation value is the average one
would obtain given many measurements of the parameter in many universes. In
Table~\ref{tab:outofbox} we have displayed the expectation values of \Omm\
and $\fvn$, together with the bestfit value of \Omm\ for comparison. In the
cases in which the bestfit value of \Omm\ takes the smallest values, namely
the Constitution, SALT2 and MLCS31 samples of \citet{Hicken09} and the Union2
sample of \citet{Amanullah10}, the bestfit value of \Omm\ differs from the
expectation value by a factor close to one standard deviation.

The extremely small bestfit values of \Omm\ for the Union and Constitution
samples --- or equivalently the unusually large values of $\fvn$ --- match
those found by~\citet{KFL}, which led these researchers to draw unfavourable
conclusions about the TS model. However, \KFL\ reasoned that this was due
to inclusion of extra data in the Union and Constitution samples, as
compared to the Riess07 sample. In particular, there are a larger number of
\sne\ in the redshift range $0.35<z<0.4$ in the Union sample, and in the
range $0<z<0.2$ in the Constitution sample. While the inclusion of \sne\
at extremely close distances below the scale of statistical homogeneity
$z<0.033$ is a separate systematic issue that needs to be carefully
investigated, the results of the MLCS2k2--reduced samples in
Table~\ref{tab:outofbox} refute the claim of \cite{KFL} that it is the
greater sample size that is the main issue\footnote{Among other general
statements \citet{KFL} make a particular comment: ``The best-fitting
parameters of the FB model are extremely sensitive to small changes in the
\sne\ data as it needs to compensate for these by a large variation in $\fvi$
when fitted to an another redshift distribution with a different amount of
error on each \sne. In addition, there is a special set of values for $\fvi$
which will mimic \LCDM\ parameters well, \dots'' These statements --
together with other statements about the role of the initial void fraction,
$\fvi$ -- are erroneous, since as we have already observed the
values of \Omm\ and $\fvn$ are insensitive to the value of $\fvi$. \citet{KFL}
appear to have not understood the quantitative role of this parameter
in the general exact solution \citep{sol}.}.
The parameters of the TS model are not sensitive to small changes in the \sne\
data, as \KFL\ maintain. Rather, we see that the primary question is the method
of data reduction. While the MLCS31 sample, fit with $\RV=3.1$ produces
results close to the SALT/SALT II fits, the MLCS17 sample, with the largest
number of \sne\ fit by the MLCS2k2 method, yields a bestfit value
$\OmMn=0.20^{+0.10}_{-0.17}$, and the SDSS-II sample of \citet{Kessler09} a
bestfit value of $\OmMn=0.40^{+0.09}_{-0.11}$. The parameters found from
the MLCS17 and SDSS-II samples agree with those of the Riess07 Gold sample
to within one standard deviation.

We conclude that the MLCS-reduced \sn\ samples, with appropriate treatment
of host galaxy reddening, provide a better fit to the TS model than the
SALT--reduced samples. We will see in Sec.~\ref{sect:shs} that this carries
through to the Bayesian statistical evidence as well. Our results
therefore support the observation of \citet{sollerman09}, who find
that for the SDSS-II supernovae of \citet{Kessler09} reduced with the MLCS2k2
fitter, nonstandard cosmological models can provide a better fit to the data
than the \LCDM\ model.

\subsection{Recalibrating the SALT \sne}\label{sect:recal}

Having established that the primary issue for the goodness of fit of the
TS model relative to the \LCDM\ model is the data reduction
method used, we should examine the extent to which implicit use of the \LCDM\
model in data calibration affects the SALT/SALT II samples. It is difficult
to assess all the ways in which the assumption of a homogeneous cosmology
is built into the SALT data reduction procedure. However, it is relatively
straightforward to adapt the \cosfit\ code, which implements the SALT
procedure, to use the luminosity distance of other cosmological models.

We replaced the module of \cosfit\ that implements the spatially
flat \LCDM\ luminosity distance by one that computes the TS
luminosity distance (\ref{eq:tslumdist}). Leaving the rest of the
\cosfit\ code unchanged, we reran the parameter fits.
This amounts to taking the published stretch and colour
parameters for each supernova, and recomputing \Omm\ along with the
empirical parameters $\scrM$, $\alpha$ and $\beta$ of eq.~(\ref{eq:salt}).

\begin{table}
\begin{center}
\begin{minipage}[t]{\linewidth}
\caption{Expectation values for the parameters of the timescape model from
SALT-reduced \sn\ data, recomputed with the timescape model luminosity
distance.
Distance normalisation is arbitrary. The bestfit value of \Omm\
is in brackets.
\label{tab:tsbestfit_recal}}
\begin{tabular}{lcccc}
\hline\hline
Dataset & $N$ & $\chi^2$ & \Omm & $f_{v0}$ \\
\hline
Union (TS) & 307 & 350.6 & $0.13^{+0.10}_{-0.08}(0.09)$ & $0.91^{+0.07}_{-0.06}$
\\ [1ex]
Union ($\Lambda$CDM) & & 344.1 & $0.28^{+0.03}_{-0.03}(0.28)$ & \\ [1ex]
Const. (TS) & 397 & 319.7 & $0.13^{+0.09}_{-0.08}(0.08)$ &
$0.91^{+0.06}_{-0.06}$ \\ [1ex]
Const. ($\Lambda$CDM) & & 312.9 & $0.29^{+0.03}_{-0.03}(0.28)$ & \\ [1ex]
\hline
\end{tabular}
\end{minipage}\hfill
\end{center}
\end{table}
The values of \Omm\ and the minimum $\chi^2$ that result from this
reanalysis are displayed in Table~\ref{tab:tsbestfit_recal}. Since the
comparable published values for the \LCDM\ model often include BAO or WMAP
priors, we also show the corresponding parameter values we obtained
ourselves using \cosfit\ applied to the spatially flat \LCDM\ model. The
results indicate that the expectation values of \Omm\ increase only very
slightly from the values of Table~\ref{tab:outofbox}. However, the bestfit
value of \Omm\ is much closer to the expectation
value for the Constitution sample. The parameters $\alpha$, $\beta$, and $M_B$
(calculated from $\scrM$ with\footnote{We will simply adopt the same
Hubble constant normalization as \citet{Hicken09}, who took this value.}
\H={65}) are shown in Table~\ref{tab:tsbestfit_globpar}. Both the TS and
\cosfit\ results are well within the uncertainties quoted in \citet{Hicken09}.
Given that the Constitution set contains the Union set, it is not surprising
that their corresponding \Omm\ values should be the same, when calculated
for each model, as Table~\ref{tab:tsbestfit_recal} shows. The addition of
the new CfA3 \sne\ to the Union sample changes the fits of the global
parameters $\alpha$, $\beta$, and $\scrM$, and consequently the estimate
of \Omm, but Table~\ref{tab:tsbestfit_globpar} shows these changes are
relatively small.
\begin{table}
\begin{center}
\begin{minipage}[t]{\linewidth}
\caption{Values for the global parameters from SALT-reduced \sn\ data,
computed by \cosfit\ with the \LCDM\ and TS luminosity distances.
\label{tab:tsbestfit_globpar}}
\begin{tabular}{lccc}
\hline\hline
Dataset & $\alpha$ & $\beta$ & $M_B$ \\
\hline
Union (TS) & $1.32^{+0.12}_{-0.12}$ & $2.37^{+0.12}_{-0.12}$ & -19.42 \\ [1ex]
Union ($\Lambda$CDM)\footnotemark[1] & $1.33^{+0.12}_{-0.12}$ &
$2.38^{+0.13}_{-0.12}$ & -19.46 \\ [1ex]
Const. (TS) & $1.29^{+0.10}_{-0.10}$ & $2.49^{+0.11}_{-0.11}$ & -19.43 \\ [1ex]
Const. ($\Lambda$CDM)\footnotemark[2] & $1.31^{+0.10}_{-0.10}$ &
$2.50^{+0.11}_{-0.11}$ & -19.46 \\ [1ex]
\hline
\end{tabular}
\end{minipage}\hfill
$^1$\citet{Kowalski08} values: $\alpha=1.24^{+0.10}_{-0.10}$,
$\beta=2.28^{+0.11}_{-0.11}$.
$^2$\citet{Hicken09} values: $\alpha=1.34^{+0.08}_{-0.08}$,
$\beta=2.59^{+0.12}_{-0.08}$, $M_B=-19.46$ with \H={65}.
\end{center}
\end{table}

\begin{figure}
\begin{center}
\caption{Differences in the distance moduli obtained by a SALT fit to the
Constitution sample using the \cosfit\ code adapted to: (i) the spatially flat
\LCDM\ model, and, alternatively, (ii) the TS model; as a function of
redshift.
\label{fig:recalconst}}
\includegraphics[scale=0.3,angle=270]{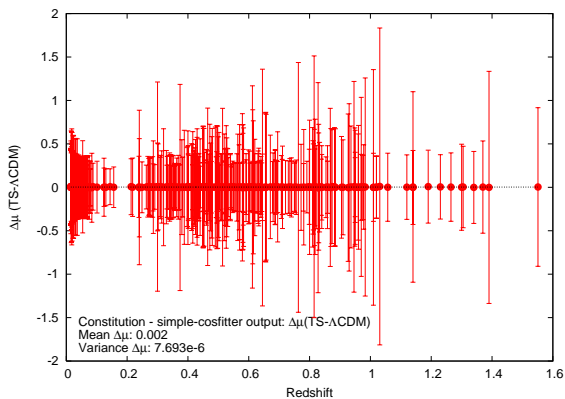}
\end{center}
\end{figure}

Overall the differences between Table~\ref{tab:outofbox} and
Table~\ref{tab:tsbestfit_recal} are relatively small. In particular, whether
or not the \LCDM\ or TS luminosity distance is assumed, the SALT-reduced
data still produces consistently higher fits to the present epoch void
fraction, $\fvn$, than that of the
MLCS-reduced samples (other than MLCS31). The resulting \Omm\ values
depend only very weakly on whether the luminosity
distance assumed by the SALT fitter is a TS one or a \LCDM\ one.
Furthermore, the distance moduli themselves show insignificant
differences: the difference between the two measures is plotted in
Fig.~\ref{fig:recalconst} (on the same scale as used in Figs.~\ref{fig:140}
and \ref{fig:140sc} for comparison).

Fig.~\ref{fig:recalconst} demonstrates that use of the TS luminosity distance
is capable of reproducing the SALT \LCDM\ results to great precision. Thus
the differences in the expectation values of cosmological parameters found
in Table~\ref{tab:outofbox} between the MLCS and SALT fitters must be
a consequence of systematic differences, rather than the luminosity distance
relation assumed by SALT.

As a check that the differences between the fitters are not simply a
consequence of the inclusion of different \sne\ subsamples, we have compared
the cosmological parameters determined from the subsample of 140 \sne\ common
to the Riess07 Gold, Union, and Constitution datasets, which was used in
generating Figs.~\ref{fig:140} and \ref{fig:140sc}. In Table~\ref{tab:140} we
show the TS cosmological parameters determined using the both the published
data for the Riess07 dataset\footnote{For the subsample of 140 \sne\ from
Riess07 gold dataset a fit of the Hubble constant gives $\Hm=$ ($61.4^{+1.4}
_{-1.5}$, $62.3^{+1.4}_{-1.8}$) $\kmsMpc$ for the TS and spatially flat \LCDM\
models respectively as compared to the respective values $\Hm=$ ($61.7^{+1.2}
_{-1.1}$, $62.6\pm1.3$) $\kmsMpc$ for the full sample of 182 \sne. As remarked
in Sec.~\ref{sect:H0} this is not an absolute determination of $\Hm$, as the
data assumes an overall calibration; but relatively speaking, the favoured
value of $\Hm$ is only very slightly reduced by restricting to the
subsample.}, Union and Constitution samples, and also our own SALT fit (with
the TS luminosity distance) of these 140 points alone. We see that the subsets
of published data values produce higher estimates of \Omm\ for each sample
than each of the complete sets given in Table~\ref{tab:outofbox}, and that
the greatest percentage increase is for the Constitution sample. Furthermore,
a \cosfit\ fit to the subsample of 140 \sne\ alone also produces higher
estimates of \Omm, which differ somewhat from the subsample of the published
values only in the case of the Union sample. However, in all cases the
SALT--reduced Union and Constitution subsamples still have a significantly
lower value of \Omm\ than the MLCS reduced Riess07 subsample. Consequently,
intrinsic differences in the MLCS and SALT methods appear to be the dominating
cause of the variance in cosmological parameter estimates for the TS model.
\begin{table}
\begin{center}
\begin{minipage}[t]{\linewidth}
\caption{Expectation values of \Omm\ (with bestfit values in
brackets), for the 140 \sne\ common to the Riess07 Gold (R), Union (U), and
Constitution (C) samples. For the Union and Constitution subsamples the
results of fits to the published data (p); and to \cosfit\ fit data (f),
are both shown. Bayes factors, with priors $0.01\le\OmMn\le0.95$ (and
$55\le\Hm\le75\kmsMpc$ where relevant), for the TS
model relative to the spatially flat \LCDM\ model are also given.
\label{tab:140}}
\begin{tabular}{lccc}
\hline\hline
Dataset & \Omm & $\fvn$ & ln $B$ \\
\hline
R140 & $0.33^{+0.15}_{-0.14}(0.36)$ & $0.76^{+0.13}_{-0.13}$ & 0.14 \\ [1ex]
U140(p) & $0.21^{+0.17}_{-0.20}(0.23)$ & $0.85^{+0.14}_{-0.13}$ & 0.43 \\ [1ex]
U140(f) & $0.16^{+0.12}_{-0.10}(0.13)$ & $0.89^{+0.07}_{-0.09}$ & 0.14 \\ [1ex]
C140(p) & $0.17^{+0.16}_{-0.16}(0.17)$ & $0.88^{+0.11}_{-0.12}$ & 0.56 \\ [1ex]
C140(f) & $0.18^{+0.12}_{-0.12}(0.17)$ & $0.87^{+0.09}_{-0.09}$ & 0.17 \\ [1ex]
\hline
\end{tabular}
\end{minipage}\hfill
\end{center}
\end{table}

The Bayes factors, representing the integrated likelihood of the TS model
over that of the spatially flat \LCDM\ model have also been given in
Table~\ref{tab:140}. By the Jeffreys scale \citep{Trotta2008}
these results indicate that the models are statistically indistinguishable
for the subsample of 140 \sne, regardless of the fitter used. For the Riess07
Gold data this is interesting, given that the whole sample of 182 \sne\ gives
$\ln B=-1.20$ with mild positive evidence in favour of the \LCDM\ model.
Although the inclusion of the additional 42 \sne\ in the Riess07 Gold sample
does not greatly affect the values of cosmological parameters, it significantly
changes the relative goodness of fit of the TS and \LCDM\ models. Since the
Constitution sample simply augments the Union sample with more recent data, the
reason for the exclusion of the 42 \sne\ in question boils down to the
differences in the selection cuts of \citet{Kowalski08} as compared to
those of \citet{Riess06}.

\subsection{The Statistical Homogeneity Scale}\label{sect:shs}

There are a huge number of potential sources of systematic errors, which,
treated differently in the MLCS and SALT fitters, might be responsible for the
differences in parameter estimates obtained for the TS model. As we have
already discussed in Sec.~\ref{sect:bubble}, in the TS model an average
Hubble flow is only expected on scales greater than the statistical
homogeneity scale\footnote{We prefer to the terminology ``statistical
homogeneity scale'' to the more commonly used ``Hubble bubble'' terminology,
since the latter is often taken to be a single large local void. In the \LCDM\
context, such a feature is, strictly speaking, anomalous. In the TS
model an {\em apparent} Hubble bubble at any typical wall location is
an expected for averages below the statistical homogeneity scale,
given the dominance of of $30/h\,$Mpc diameter voids by volume in the late
epoch universe.} (SHS). Thus for consistency, in performing any parameter
fits on the TS model, a cut should be applied to data at redshifts below
the SHS.

For the SALT datasets the inclusion of significant numbers of \sne\ below
the SHS could conceivably bias the global fits of $\alpha$, $\beta$, $\scrM$
and \Omm. The potential impact of the SHS on the MLCS datasets is less
clear, as the fitter is trained using a set of nearby \sne\ that are far
enough into the Hubble flow for peculiar velocities to be negligible, yet
still close enough for the linear Hubble law to hold \citep{JRK}. In practise,
the training set includes some \sne\ within the SHS, so there may be
systematic issues associated with the SHS, just much more subtle.

\begin{table}
\begin{center}
\begin{minipage}[t]{\linewidth}
\caption{Parameter values for \sn\ datasets, applying homogeneity scale cuts,
the first at the Hubble bubble radius of $\zmin=0.024$~(e.g. \citet{JRK}), the
second at $\zmin=0.033$, corresponding to the scale of statistical homogeneity
estimated to be $\sim100~h^{-1}$ Mpc. Expectation values of \Omm\ are shown,
with bestfit values in brackets. For the SALT/SALT--II fits (Union,
Constitution, SALT2, Union2) the parameters have been recomputed by
adapting \cosfit\ to the TS model in each case.
\label{tab:hubblebubble}}
\begin{tabular}{lccccl}
\hline\hline
Dataset & $z$ cut & $N$ & $\chi^2$ & \Omm & $\ln B$\\
\hline
\multirow{3}{*}{Gold} & $\ge0.024$ & 182 & 162.7 & $0.30^{+0.14}_{-0.13}(0.33)$
& -1.20
\\ [1ex]
& $\ge0.033$ & 169 & 151.8 & $0.31^{+0.15}_{-0.13}(0.34)$ & -1.18 \\ [1ex]
\hline
\multirow{3}{*}{R140} & $\ge0.024$ & 140 & 102.7 &
$0.33^{+0.16}_{-0.14}(0.36)$ &
0.14
\\ [1ex]
& $\ge0.033$ & 132 & 96.2 & $0.26^{+0.20}_{-0.25}(0.30)$ & 0.78 \\ [1ex]
\hline
\multirow{3}{*}{MLCS17} & None & 372 & 401.7 & $0.18^{+0.13}_{-0.15}(0.20)$
& 0.77 \\ [1ex]
& $\ge0.024$ & 282 & 315.7 & $0.17^{+0.13}_{-0.16}(0.19)$ & 0.37 \\ [1ex]
& $\ge0.033$ & 234 & 260.2 & $0.19^{+0.14}_{-0.17}(0.21)$ & 0.57 \\ [1ex]
\hline
\multirow{3}{*}{MLCS31} & None & 366 & 429.5 & $0.07^{+0.05}_{-0.06}(0.01)$
& -1.57 \\ [1ex]
& $\ge0.024$ & 278 & 332.2 & $0.09^{+0.08}_{-0.08}(0.01)$ & 0.13 \\ [1ex]
& $\ge0.033$ & 229 & 263.3 & $0.11^{+0.11}_{-0.10}(0.03)$ & 1.09 \\ [1ex]
\hline
\multirow{3}{*}{SDSS-II} & None & 288 & 240.8 & $0.39^{+0.11}_{-0.09}(0.40)$
& 0.09 \\ [1ex]
& $\ge0.024$ & 284 & 238.4 & $0.40^{+0.11}_{-0.10}(0.41)$ & 0.27 \\ [1ex]
& $\ge0.033$ & 272 & 214.5 & $0.42^{+0.10}_{-0.10}(0.44)$ & 0.53 \\ [1ex]
\hline
\multirow{3}{*}{Union} & None & 307 & 350.6 & $0.13^{+0.10}_{-0.08}(0.07)$ &
-2.04 \\ [1ex]
& $\ge0.024$ & 288 & 333.4 & $0.15^{+0.10}_{-0.09}(0.14)$ & -1.53 \\ [1ex]
& $\ge0.033$ & 275 & 318.0 & $0.18^{+0.11}_{-0.11}(0.20)$ & -0.86 \\ [1ex]
\hline
\multirow{3}{*}{Const.} & None & 397 & 319.6 &
$0.13^{+0.09}_{-0.08}(0.06)$
& -1.54 \\ [1ex]
& $\ge0.024$ & 351 & 293.8 & $0.13^{+0.09}_{-0.08}(0.09)$ & -1.57 \\ [1ex]
& $\ge0.033$ & 309 & 275.9 & $0.16^{+0.10}_{-0.10}(0.15)$ & -1.06 \\ [1ex]
\hline
\multirow{3}{*}{SALT2} & None & 351 & 402.5 &
$0.10^{+0.08}_{-0.06}(0.02)$
& -2.25 \\ [1ex]
& $\ge0.024$ & 278 & 342.1 & $0.11^{+0.08}_{-0.06}(0.03)$ & -2.22 \\ [1ex]
& $\ge0.033$ & 235 & 305.5 & $0.13^{+0.09}_{-0.07}(0.09)$ & -1.55 \\ [1ex]
\hline
\multirow{3}{*}{Union2} & None & 557 & 520.3 &
$0.09^{+0.07}_{-0.08}(0.05)$ & -2.65 \\ [1ex]
& $\ge0.024$ & 504 & 483.5 & $0.10^{+0.08}_{-0.06}(0.09)$ & -2.25 \\ [1ex]
& $\ge0.033$ & 457 & 428.4 & $0.10^{+0.08}_{-0.06}(0.15)$ & -3.46 \\ [1ex]
\hline
\multirow{3}{*}{CSP} & None & 56 & 62.3 &
$0.11^{+0.08}_{-0.10}(0.01)$ & -4.23 \\ [1ex]
& $\ge0.024$ & 47 & 69.1 & $0.12^{+0.13}_{-0.11}(0.01)$ & -4.03 \\ [1ex]
& $\ge0.033$ & 43 & 46.0 & $0.13^{+0.12}_{-0.12}(0.01)$ & -3.34 \\ [1ex]
\hline
\end{tabular}
\end{minipage}\hfill
\end{center}
\end{table}
As a first check on whether the discrepancy between the values of \Omm\
determined by the SALT and MLCS methods can be accounted for by making
cuts at the SHS scale, we have determined parameters for fits to the TS
model with cuts made: (i) by excluding objects at redshifts $z<0.024$,
which corresponds to the $\Hm d_{\rmn{SN}}\simeq7400\,$km$\,$sec$^{-1}$
Hubble bubble partition assumed by \citet{JRK,Riess06} and
\citet{Hicken09}); and (ii) by excluding objects with
redshifts $z<0.033$, corresponding to the estimated scale of statistical
homogeneity, $100/h$ Mpc. The resulting values of \Omm\
are compared with the values obtained from the full dataset in
Table~\ref{tab:hubblebubble}. Bayes factors for a fit of the TS model
relative to the spatially flat \LCDM\ model are also displayed.

In all cases the relevant cuts lead to somewhat larger values of \Omm,
with the exception of the subsample of 140 \sne\ from the Riess07 Gold
dataset, which was discussed in Table~\ref{tab:140}. In the SALT
cases the increase in the expectation value of \Omm\ is not particularly
large. However, the bestfit value of \Omm\ increases generally by a factor
of three from the full sample to the SHS cut sample. Since the bestfit value
of \Omm\ is much closer to the expectation value for the SALT samples
once SHS cuts are made, it shows that the fits are no longer so strongly
skewed away from being Gaussian with bestfit values of \Omm\ at unreasonably
small values. This demonstrates that the main criticism of \citet{KFL}
is invalid once the SHS cut is made.

In all the MLCS2k2 cases, the Bayes factor improves in favour of the TS model
with the redshift cuts, but the improvement is weak and, with the exception
of MLCS31, generally not enough to statistically distinguish the
models. For the SALT reduced data, by contrast, the data
generally indicate mild positive evidence against the TS model on the Jeffreys
scale~\citep{Trotta2008}. For all SALT datasets apart from Union2 the Bayesian
evidence in favour of the \LCDM\ model is weaker once the SHS cut is applied.
\begin{figure}
\caption{Bestfit values (solid line) and expectation values (dotted line)
of \Omm\ for successive redshift cuts for eight \sne\ samples. The
probability distributions for SALT \Omm\ fits (left column) make a
transition from negative skew to positive skew, while those MLCS samples
(right column) which already provide better fits to the TS model are always
positively skewed. \label{fig:omplots}}
\begin{flushright}
\includegraphics[width=0.5\textwidth]{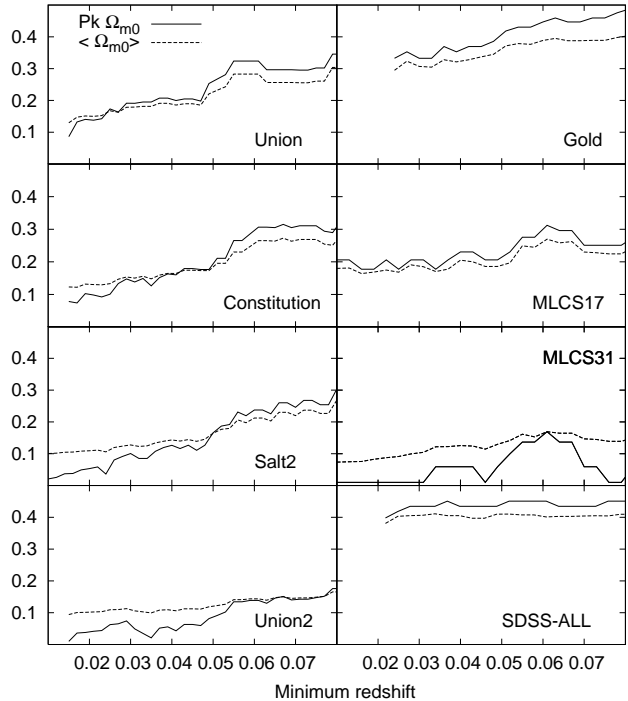}
\end{flushright}
\end{figure}

\begin{figure*}
\caption{Bayesian evidence: $\ln B$ for the TS model relative to the
spatially flat \LCDM model for successive redshift cuts for nine \sne\
samples. The SALT/SALT II reduced sets are in the left panel, the MLCS2k2
reduced samples in the right panel. \label{fig:Bayesplots}}
\begin{center}
\includegraphics[angle=270,width=0.9\textwidth]{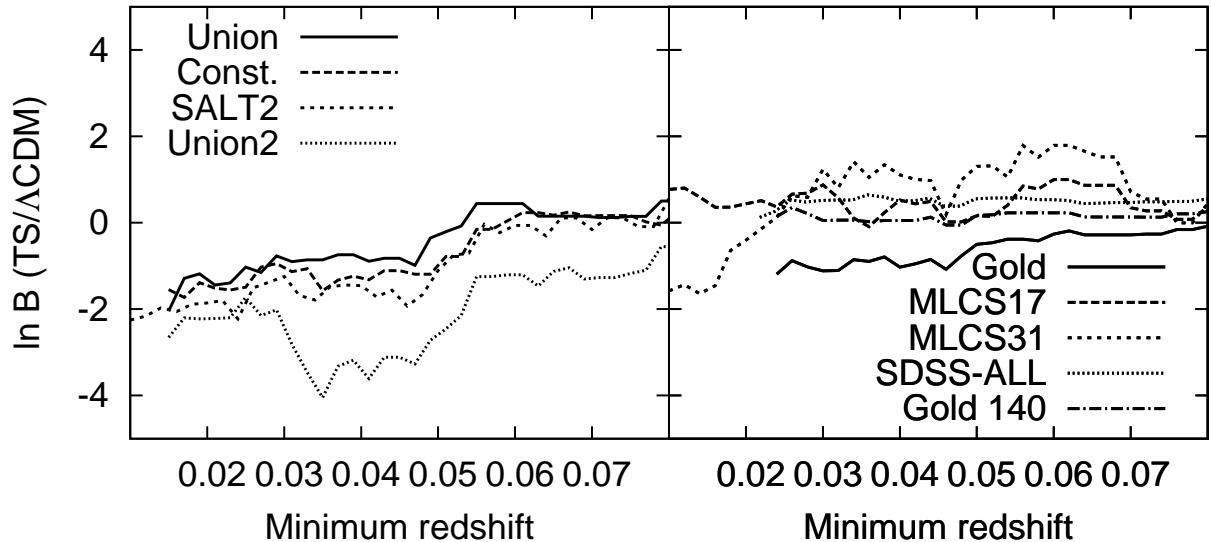}
\end{center}
\end{figure*}

In order to further test the issue of redshift cuts, we have repeated
our analysis by applying a cut at redshifts which range from the minimum
value in each dataset up to $\zmin=0.08$. The results are shown in
Figs.~\ref{fig:omplots} and Figs.~\ref{fig:Bayesplots}. The Bayes factor for
the SHS ($\zmin= 0.033$) cut on the Union2 sample turns out to mark the
beginning of a downward trough in an otherwise increasing trend.

Fig.~\ref{fig:omplots} provides a direct demonstration of how excluding
the \sne\ below the SHS leads to to a better agreement between the bestfit
and expectation values of \Omm\ in the SALT/SALT II reduced datasets. In
fact, it also reveals differences between the SALT and SALT II fitters. For
the Union and Constitution datasets, reduced by SALT, we see that below the
SHS the bestfit value is below the expectation value, i.e., negatively skewed.
For cuts in the range $0.03\lsim\zmin\lsim0.05$ the bestfit and expectation
values of \Omm\ are roughly comparable for these samples, while for
$\zmin\gsim0.05$ the bestfit value is positively skewed. For SALT2 and Union2,
reduced by SALT II, the bestfit value remains negatively skewed for cuts up to
$\zmin\lsim0.055$.

In the case of the datasets that the TS model fits well -- Riess07 Gold,
MLCS17 and SDSS-II -- the skew is always positive, i.e., the bestfit value is
greater than the expectation value. The skew is larger for the cuts
at higher redshifts, particularly for the Gold and SDSS-II samples. For MLCS31,
the bestfit value is negatively skewed, and the fit is generally poor, in the
sense that this peak probability is very significantly skewed.

These results suggest that the way that the SALT fitters treat nearby
objects -- the inclusion of large numbers of them as well as the treatment of
their color variations and host galaxy dust -- does affect the TS
cosmology fits.

Finally, for completeness we have added the recently published first
dataset from the Carnegie Supernova Project (CSP) \citep{Freedman09} to
Table~\ref{tab:hubblebubble}. This is a much smaller dataset than the others.
The CSP differs from other projects by working in the near infrared, and
consequently the data is differently reduced. However, in analysing their
data \citet{Freedman09} have adopted a reduced--$\chi^2=1$ approach,
conceptually similar to the approaches used in the SALT/SALT-II fits.
Nuisance parameters, including $\RV$, are determined by minimising Hubble
residuals for the whole diagram. In Table~\ref{tab:hubblebubble} we have
only presented fits to the TS model for the ``out of the box'' data of
\cite{Freedman09}, which was fit to a FLRW model with constant dark energy
equation of state parameter, $w$. To be more confident about the conclusions
we should redo the data reduction for the TS model. However, such a task
requires considerable effort, and we will defer this until such a time as
significantly more data is available. At this stage we note that
\citet{Freedman09} found a value of $\RV=1.74\pm0.27\,$(stat)$\pm0.01\,$(sys),
considerably lower than the Milky Way $\RV=3.1$ value, which is consistent
with the values of the corresponding values of $\beta$ typical in the
SALT/SALT-II fits. Furthermore, the very low expectation values of \Omm\ are
most consistent with SALT2 and Union2. While the bestfit value of \Omm\
is essentially driven to zero, giving an extremely skewed distribution,
we must recall that this was also the case for Union2 ``out of the box''
data in Table~\ref{tab:outofbox} before the SALT-II parameters were
recomputed for the TS luminosity distance. Given the similarity in
approach, we would expect an improvement in the bestfit values if the TS
luminosity distance was assumed in the data reduction.

\subsection{Systematic issues for MLCS}\label{sect:mlcssys}
The erratic results from MLCS31 ($\RV=3.1$) for the minimum redshift
cuts in Figs.~\ref{fig:omplots} and Figs.~\ref{fig:Bayesplots}, and the
relatively stable results from MLCS17 ($\RV=1.7$) and SDSS-II ($\RV=2.18$)
suggest that while the MLCS samples are better fit by the TS model, such fits
are still highly affected by the treatment of dust extinction and reddening.
However, the issues are much subtler than simply the value of $\RV$ assumed.
In particular, the Riess07 gold dataset \citep{Riess06} and the MLCS31
dataset \citep{Hicken09} are both fit with $\RV=3.1$ and yet they produce
expectation values of \Omm\ which differ by a factor of three, once the
SHS cut is made. Similarly, the MLCS17 dataset \citep{Hicken09} and
SDSS-II dataset are both fit with low $\RV$ values, respectively 1.7 and
2.18, while producing expectation values of \Omm\ which differ by a factor of
two, for the SHS-cut samples.

These large differences in the probable values of \Omm\ suggest that the
different assumptions in sample selection made with different versions of
MLCS2k2 have a very significant impact in the determination of TS model
parameters. As a check, we have redone the analysis of Table~%
\ref{tab:hubblebubble} for the following samples: Sample A of the 76 \sne\
common to the Riess07 Gold and MLCS31 datasets; and Sample B of the 74 \sne\
common to the Riess07 Gold and MLCS17 datasets. The two samples include the
same \sne\ apart from four objects -- three \sne, SN 1999gp at $z=0.026$,
SN 1991U at $z=0.033$, and SN 1992J at $z=0.046$ are in Gold and MLCS31 but not
in MLCS17, and one, SN 2004D4dw at $z=0.961$, is in Gold and MLCS17 but not in
MLCS31. We have tested the both samples in full, and with a SHS cut. Of the
four \sne\ not common to both Sample A and Sample B, only one, SN 1999gp is
below the SHS, though SN 1991U at $z=0.033$ is borderline. In addition, we
make a final cut to include the 66 \sne\ above the SHS which are common to
both samples.
\begin{table}
\begin{center}
\begin{minipage}[t]{\linewidth}
\caption{Parameter values for \sn\ datasets, applying homogeneity scale cuts,
to the 76 \sne\ common to Riess07 Gold and MLCS31 (Sample A); and to the 74
\sne\ common to Riess07 Gold and MLCS17 (Sample B).
\label{tab:sampleAB}}
\begin{tabular}{lccccl}
\hline\hline
Dataset & $z$ cut & $N$ & $\chi^2$ & \Omm & $\ln B$\\
\hline
\multirow{3}{*}{A Gold} & $\ge0.024$ & 76 & 60.7 & $0.20^{+0.20}_{-0.19}(0.13)$
& 0.73
\\ [1ex]
& $\ge0.033$ & 68 & 51.0 & $0.19^{+0.19}_{-0.18}(0.02)$ & 0.41 \\ [1ex]
& $\ge0.033$ & 66 & 41.3 & $0.23^{+0.24}_{-0.22}(0.23)$ & 0.84 \\ [1ex]
\hline
\multirow{3}{*}{A MLCS31} & $\ge0.024$ & 76 & 76.6 &
$0.14^{+0.13}_{-0.13}(0.01)$ & 0.09\\ [1ex]
& $\ge0.033$ & 68 & 65.7 & $0.13^{+0.11}_{-0.12}(0.01)$ & -0.73 \\ [1ex]
& $\ge0.033$ & 66 & 51.9 & $0.16^{+0.16}_{-0.15}(0.01)$ & 0.40 \\ [1ex]
\hline
\multirow{2}{*}{B Gold} & $\ge0.024$ & 74 & 50.5 & $0.24^{+0.23}_{-0.23}(0.26)$
& 0.80
\\ [1ex]
& $\ge0.033$ & 67 & 41.3 & $0.23^{+0.23}_{-0.22}(0.23)$ & 0.93 \\ [1ex]
\hline
\multirow{2}{*}{B MLCS17}& $\ge0.024$ & 76 & 76.2
& $0.18^{+0.20}_{-0.17}(0.11)$ & 0.93 \\ [1ex]
& $\ge0.033$ & 67 & 74.5 & $0.18^{+0.19}_{-0.17}(0.01)$ & 0.86 \\ [1ex]
\hline
\end{tabular}
\end{minipage}\hfill
\end{center}
\end{table}

From Table~\ref{tab:sampleAB} we see that the expected and bestfit
values of \Omm\ for the Riess07 Gold sample are somewhat reduced relative to
those of the full sample given in Table~\ref{tab:sampleAB}, and that they
agree with the corresponding values for the MLCS31 and MLCS17 subsamples
within the 1$\si$ uncertainties. Nonetheless there are still some differences
in the central expectation values, particularly in the case of the MLCS31
sample. Since the Gold and MLCS31 samples assume the same $\RV=3.1$,
the differences might be a consequence of the different assumptions
about extinction priors \citep{JRK} made in the different implementations of
MLCS2k2.

While in most cases the Bayesian evidence for the TS model
relative to the spatially flat \LCDM\ model is improved for the subsamples
of Table~\ref{tab:sampleAB}, the most striking change comes about when the
MLCS17-excluded \sne\ SN 1991U and SN 1992J are excluded from the SHS--cut
Gold and MLCS31 A samples, reducing these from 68 to 66 objects. The effect
of this is to substantially increase \Omm\ in both cases, to remove a
dramatic negative skew of the bestfit \Omm\ in the Gold subsample, and
to change the Bayesian evidence by an order of magnitude, (a factor 3 in
$B$), in the case of the MLCS31 subsample. We have not investigated
the reasons for the exclusion of SN 1991U and SN 1992J in the MLCS17 sample;
however, they do indeed appear to be outliers in the present analysis.
To test this hypothesis, we have recomputed the Riess07 Gold and MLCS31
entries of Table~\ref{tab:hubblebubble} with the three MLCS17-excluded
\sne\ removed. The effect of removing two \sne\ on these larger samples
might expected to be less. Nonetheless, the SHS-cut samples do show
a dramatic change. The Bayesian evidence for the TS model is significantly
increased in each case.

\begin{table}
\begin{center}
\begin{minipage}[t]{\linewidth}
\caption{Recalculation of Table~\ref{tab:hubblebubble} for the \citet{Riess06}
Gold sample and the \citet{Hicken09} MLCS31 samples, with three low redshift
\sne\ that were excluded by \citet{Hicken09} from their MLCS17 sample,
SN 1991U, SN 1992J and SN 1999gp, excluded.
\label{tab:cutmlcs17}}
\begin{tabular}{lccccl}
\hline\hline
Dataset & $z$ cut & $N$ & $\chi^2$ & \Omm & $\ln B$\\
\hline
\multirow{2}{*}{Gold} & $\ge0.024$ & 179 & 151.0 & $0.34^{+0.13}_{-0.11}(0.36)$
& -0.77
\\ [1ex]
& $\ge0.033$ & 167 & 139.9 & $0.31^{+0.15}_{-0.13}(0.34)$ & -0.63 \\ [1ex]
\hline
\multirow{3}{*}{MLCS31} & None & 363 & 414.9 &
$0.08^{+0.05}_{-0.07}(0.01)$ & -1.27\\ [1ex]
& $\ge0.024$ & 276 & 317.2 & $0.10^{+0.08}_{-0.09}(0.01)$ & 0.50 \\ [1ex]
& $\ge0.033$ & 227 & 248.4 & $0.13^{+0.13}_{-0.12}(0.08)$ & 1.68 \\ [1ex]
\hline
\end{tabular}
\end{minipage}\hfill
\end{center}
\end{table}

Finally, given that individual \sne\ which were excluded from either
the MLCS17 or MLCS31 samples by \citet{Hicken09} are potential outliers,
(at least for the \LCDM\ model), we recompute
Table~\ref{tab:hubblebubble} for the 352 \sne\ common to both MLCS31
and MLCS17 samples, using each $\RV$ normalization. The results are
shown in Table~\ref{tab:mlcscommon}. For the SHS-cut samples, there
is no overall change to the \Omm\ parameter for the MLCS17 sample,
but the bestfit value of MLCS31 sample is increased to be more in
line with the Table~\ref{tab:cutmlcs17} result. There is now positive
Bayesian evidence for the TS model versus the spatially flat \LCDM\ model
for MLCS17 as well as MLCS31 with the SHS cut.
\begin{table}
\begin{center}
\begin{minipage}[t]{\linewidth}
\caption{Recalculation of Table~\ref{tab:hubblebubble} for the 352 \sne\
common to both the MLCS31 and MLCS17 samples of \citet{Hicken09}.
\label{tab:mlcscommon}}
\begin{tabular}{lccccl}
\hline\hline
Dataset & $z$ cut & $N$ & $\chi^2$ & \Omm & $\ln B$\\
\hline
\multirow{3}{*}{MLCS17} & None & 352 & 366.9 &
$0.16^{+0.13}_{-0.15}(0.17)$ & 1.20\\ [1ex]
& $\ge0.024$ & 266 & 293.8 & $0.16^{+0.14}_{-0.15}(0.18)$ & 1.10 \\ [1ex]
& $\ge0.033$ & 219 & 238.1 & $0.19^{+0.14}_{-0.18}(0.21)$ & 1.37 \\ [1ex]
\hline
\multirow{3}{*}{MLCS31} & None & 352 & 403.4 &
$0.08^{+0.05}_{-0.07}(0.01)$ & -1.31\\ [1ex]
& $\ge0.024$ & 266 & 310.0 & $0.09^{+0.09}_{-0.08}(0.01)$ & 0.39 \\ [1ex]
& $\ge0.033$ & 219 & 242.1 & $0.12^{+0.12}_{-0.11}(0.07)$ & 1.55 \\ [1ex]
\hline
\end{tabular}
\end{minipage}\hfill
\end{center}
\end{table}

While the analysis of this section has not been able to resolve the
issue of how similar $\RV$ values can lead to quite different expectation
values of \Omm, as evidenced by the Riess07 Gold sample versus the MLCS31
sample, or by the SDSS-II sample versus the MLCS17 sample, it does show
that the question of extinction priors, and the \sne\ excluded by
particular priors, may be crucial to this. Ideally one should re-evaluate
the MLCS2k2 parameter fitting using the TS model at the outset.

\subsection{Parameter sensitivity in the timescape model}

The parameter \Omm\ is much more sensitive to the method of data reduction
in the case of the TS model than is the case for the \LCDM\ model. Of course,
we have not explored the goodness of fit of the \LCDM\ model when cuts are made
at the SHS in the same way that we have for the TS model, as there is no
theoretical rationale for doing so.

\begin{figure}
\begin{center}
\caption{Confidence limits for the TS model fits to $z\ge0.033$ cut samples
of Gold07 (Table \ref{tab:cutmlcs17}), SDSS-II (Table \ref{tab:hubblebubble}),
MLCS17 and MLCS31 (Table \ref{tab:mlcscommon}). In each case an overall
normalization of the Hubble constant from the published dataset is assumed.
\label{fig:mlcsall}}
\includegraphics[scale=0.48]{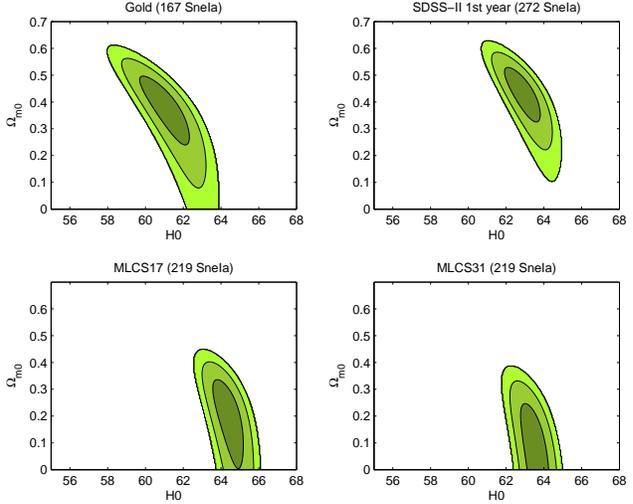}
\end{center}
\end{figure}

\begin{figure}
\begin{center}
\caption{Confidence limits for the \LCDM\ model fits to $z\ge0.033$ cut samples
of Gold07 (Table \ref{tab:cutmlcs17}), SDSS-II (Table \ref{tab:hubblebubble}),
MLCS17 and MLCS31 (Table \ref{tab:mlcscommon}). In each case an overall
normalization of the Hubble constant from the published dataset is assumed.
\label{fig:lcdmall}}
\includegraphics[scale=0.48]{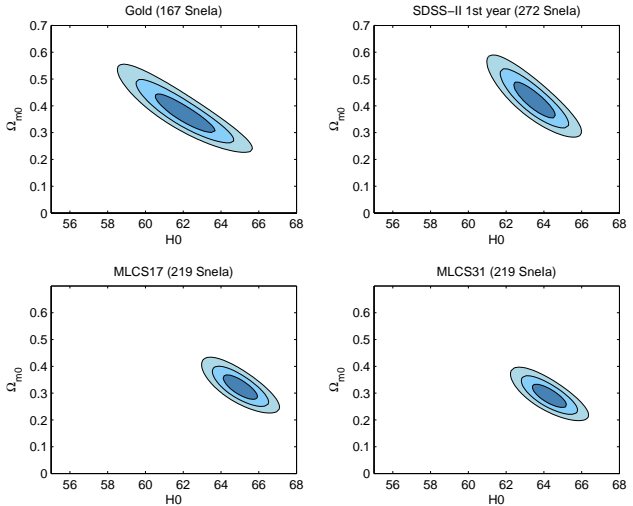}
\end{center}
\end{figure}

In order to understand differences in the degree of sensitivity of parameter
estimation between the two models, let us consider the fit of the MLCS2k2 data
once a SHS cut is made. In Fig.~\ref{fig:mlcsall} we display confidence
contours for the SHS ($z\ge0.033$) cut Gold sample of Table
\ref{tab:cutmlcs17}, the SDSS-II sample of Table \ref{tab:hubblebubble}, and
the MLCS17 and MLCS31 samples of Table \ref{tab:mlcscommon}. (In all cases
\sne\ events excluded from either the MLCS17 or MLCS31 samples have been
cut.) Corresponding plots for the spatially flat \LCDM\ model are shown in
Fig.~\ref{fig:lcdmall}. Equivalent contour plots for SALT-reduced samples are
not shown, since the value of $\Hm$ is marginalised over in the SALT data
reduction process.

Clearly there are some differences which are intrinsic to the data reduction
method, since even in the \LCDM\ model the best fit value of the matter density
parameter varies from $0.40\pm0.04$ for the SDSS-II sample to $0.26\pm0.03$
for the MLCS31 sample. The fact that the dressed matter density parameter of
the TS model is more sensitive than the corresponding value of \Omm\ for the
\LCDM\ model is essentially a consequence of different manner in which the
parameters \Omm\ and $\Hm$ affect the luminosity distances in the two models.
In the case of the \LCDM\ model the confidence contours are at roughly
$45^\circ$ for the scale chosen in Fig.~\ref{fig:lcdmall}. By contrast,
although the area of the contours is comparable in both models, on the same
scale the confidence contours for the TS model are close to vertical,
especially for the smaller values of $\OmMn\sim0.3$. Thus in the case of the
TS model the parameter $\Hm$ is more tightly constrained than in the \LCDM\
model, given whatever overall normalization of absolute magnitudes is assumed
in the data, while the parameter \Omm\ is less constrained.

Physically, these differences can be understood to be a consequence of
the fact that on one hand the \LCDM\ model has actual accelerating expansion,
whereas the apparent acceleration in the TS model is less pronounced and
its luminosity distance is closer to that of an empty Milne universe.

\section{Discussion}\label{systematics}

Even with a SHS cut there remain considerable systematic issues -- probably
concerning extinction and reddening by dust, and intrinsic \sne\ colour
variations -- which need to be resolved before one can draw any reliable
conclusions about the goodness of fit of one cosmological model over another.
These issues have been discussed by many other groups -- see, e.g.,
\citet{Hicken09,Freedman09,Kessler09,Sullivan10,Lampeitl10}.

\subsection{Reddening by dust}
There are at least four possible sources of dust: (i) dust in the Milky
Way; (ii) dust in the host galaxy; (iii) dust in the local circumstellar
environment of the supernova event; and (iv) dust in the intergalactic
medium. Whereas Milky way dust, with the reddening law $\RV=3.1$ is
well understood, we do not have direct {\rm a priori} knowledge of the
other three possible sources of dust. The present status of our
understanding is that while there appears to be no direct evidence of
an intergalactic gray dust which would significantly alter the \sne\
luminosity distance relations, the situation regarding dust within the
host galaxy, and within the supernova local environment, is a complex
one.

\subsubsection{Host galaxy dust}\label{host}
The nature of dust in the host galaxy would appear to be the most
significant systematic unknown, given that assuming different values of
$\RV$ can lead to substantial differences in the MLCS2k2 fitters, and
similarly for the value of $\beta$ in the SALT/SALT-II fitters.
Superficially, the results of Table~\ref{tab:mlcscommon} might appear
to favour the MLCS17 over the MLCS31 sample for the TS model, given the
lower value of $\chi^2$. However, it is noticeable that the results for
the MLCS17 sample show essentially no change with the cuts made up to the
SHS. By contrast, the Bayesian evidence for the TS model over the
spatially flat \LCDM\ model shows a marked improvement in Tables~%
\ref{tab:cutmlcs17} and \ref{tab:mlcscommon} when the SHS cut is
applied.

For consistency of the TS scenario an apparent Hubble bubble feature
should be seen below the SHS. Consequently the most likely expectation
for reddening and extinction by dust consistent with the above results
is that, {\em at least at low redshifts}, the reddening law for dust in
other galaxies is close to the $\RV=3.1$ law of the Milky Way.
Independent support for such a conclusion is provided by the recent studies
of \citet{Finkelman08,Finkelman10} and \citet{Folatelli10} (CSP).

\citet{Finkelman08,Finkelman10} studied dust lanes in 15 E/S0 galaxies with
$z<0.033$, and determined extinction properties by fitting model galaxies to
the unextinguished parts of the images in each of six spectral bands, and
then subtracting these from the actual images. They found an average
value $\RV=2.82\pm0.38$ for 8 galaxies in their first study
\citep{Finkelman08}, and $\RV=2.71\pm0.43$ for 7 galaxies in their second
investigation \citep{Finkelman10}. These values are a little lower than
the Milky Way value but consistent with it within the uncertainties.

\citet{Folatelli10} investigated the reddening law properties of
\sne\ in host galaxies at $z<0.08$ using well-sampled, high-precision
optical and near-infrared light curves. Although a value of $\RV\simeq1.7$
was obtained for the whole sample, once two very highly reddened
objects SN 2005A and SN 2006X were excluded, a value of $\RV\simeq3.2$,
similar to the Milky Way one, was obtained by comparison of colour excesses.
In contrast to the results obtained by comparison of colour excesses
\citet{Folatelli10} found that fits of absolute magnitude gave
$\RV\simeq1$--$2$, even when the two highly reddened \sne\ were excluded. This
discrepancy suggests that in addition to the normal interstellar reddening
produced in host galaxies, there is an intrinsic dispersion in the colours
of normal \sne\ which is correlated with luminosity but independent of the
decline rate. This would suggest that the low $\RV$ values inferred
in the MLCS17 and SDSS-II analyses may be anomalous, and furthermore
that there may be intrinsic systematic problems with the empirical methodology
assumed by the SALT/SALT-II fitters.

\subsubsection{Supernova circumstellar dust}
The actual picture may be further complicated, however, if the local
circumstellar environment of individual \sne\ is important for a
significant subclass of events. That this is potentially the case was borne
out by a recent analysis of \citet{Wang09}. They found that within a sample
of 158 relatively normal \sne, roughly one third of the objects displayed
high photospheric velocities, as determined from Si II $\lambda$6355
absorption lines. The high velocity sample of \sne\ were found to be on
average $\goesas0.1\,$mag redder than the larger group of ``normal''
supernovae. This high velocity sample includes the two very highly reddened
objects SN 2005A and SN 2006X, whose exclusion\footnote{Of the events in common
to the analysis of \citet{Folatelli10} and \citet{Wang09} all apart from three
objects SN 2004ef, SN 2005A and SN 2006X are classified by \citet{Wang09} as
``normal''. \citet{Wang09} note that there is no sharp division between their
normal and high velocity groupings when the photospheric velocity approaches a
lower value, so that blending can occur to some extent. The object SN 2004ef
is the only one treated as normal by \citet{Folatelli10} while being placed
in the high velocity sample by \citet{Wang09}.} led to the $\RV\simeq3.2$
estimate of \citet{Folatelli10}. This could either mean that the high
velocity sample have intrinsically red $B\,$--$\,V$ colours or that
they are associated with dusty local environments. Evidence for the second
possibility is directly seen in the case of SN 2006X in the nearby Virgo
cluster spiral galaxy\footnote{The circumstellar material around SN 2006X was
identified by the presence of time-variable and blue-shifted Na I D features by
\citet{Patat07}, and from a light echo by \citet{Wang08}. Spectroscopic and
photometric analysis of extinction due to circumstellar dust around SN 2006X
has been parameterised with $\RV=1.48\pm0.06$ \citep*{Wang08a}. VLT
spectropolarimetry \citep{Patat09} provides independent confirmation that
the intervening dust is different in nature from typical Milky Way dust.
SN 2006X is an unusual \sn, however, having one of the highest expansion
velocities ever observed, as well as being very highly reddened. A further
sample of 31 \sne\ has been studied for the presence of the same Na I D
features as SN 2006X by \cite{Blondin09}. The only object in their sample
other than SN 2006X which exhibited such features was the highly reddened
SN 1999cl, which is classified as a ``high velocity'' object by \citet{Wang09}.
There are 24 objects in common to the studies of \citet{Wang09} and
\cite{Blondin09} including SN 2006X and SN 1999cl. Of the 22 objects which
do not exhibit variable Na I D features, 17 are classed as ``normal'' by
\citet{Wang09} and 5 as ``high velocity''. This suggests Na I D variability
may not be systematically related to nonstandard dust.} M100 
\citep{Patat07,Wang08a,Wang08}. A model with multiple scattering of photons by
circumstellar dust is found to steepen the effective extinction law
\citep{Goobar08}. \citet{Wang09} found that their reddened high velocity
subsample preferred a lower extinction ratio $\RV\simeq1.6$ as
compared to $\RV\simeq2.4$ for the normal group, which is consistent
with this theoretical model. The difference of the $\RV$ value of the
normal group from the corresponding value of \citet{Folatelli10} may be
a further hint of a possible intrinsic colour dispersion of the normal
group.

If the sample of \citet{Wang09} is representative then perhaps of
order one third of \sne\ could have a different effective $\RV$. We note,
for example, that 78 objects are common\footnote{These objects are all at low
redshifts, $z<0.06$, and some 86\% are within the SHS.} to the sample
of \citet{Wang09} and the MLCS17 sample \citep{Hicken09}. Of these 27 are
classified as ``high velocity'' by \citet{Wang09}, and the rest as ``normal'',
which are the same rough proportions as the full sample. However,
one must take great care in making generalizations based on low redshift
samples, as low redshift \sne\ are often discovered by targeting galaxies
of particular types, such as massive galaxies. This could introduce
significant statistical biases as compared to \sne\ sampled at higher
redshifts.

Furthermore, much needs to be done to understand the astrophysics of
nonstandard circumstellar dust, as the relative
proportion of objects could be affected by evolution. For example,
it has been suggested that the nonstandard dust of SN 2006X might be
due to circumstellar material accreted from a companion star in the
red giant phase \citep{Patat07}. If the nature of the companion star
to a \sn\ progenitor is important in characterizing nonstandard dust, then
the relative statistics of such events may change with redshift.

\subsubsection{Intergalactic dust}
The possible cumulative effects of intergalactic dust ejected from galaxies
has been investigated by \citet{Menard10a}, who analysed the reddening of
$\goesas85,000$ quasars at $z>1$ due to the extended halos of 20 million
SDSS galaxies at $z\sim0.3$. They found that on large scales dust extinction
has a wavelength dependence described by\footnote{The analysis of
\citet*{Menard10b} assumed the slightly lower value $\RV=3.9\pm2.6$.}
$\RV\simeq4.9\pm3.2$. The cumulative presence of intergalactic dust along
the line of sight turns out not to affect the colour-magnitude-stretch scaling
relations, but does bias cosmological parameters in the standard cosmology
at a level comparable to current statistical errors, i.e., a few percent
\citep{Menard10b}. Accounting for the intergalactic dust led to a 6\%
increase in \Omm\ for the spatially flat \LCDM\ model. Given the increased
sensitivity of \Omm\ in the TS model to changes in the treatment of \sne\
systematics, this bias may have an even greater impact, and should be
fully investigated.

\subsection{Intrinsic colour variations}

As discussed in Sec.~\ref{host} above the analysis of \citet{Folatelli10}
suggests that there may be an intrinsic dispersion in the colours
of normal \sne\ which is correlated with luminosity but independent of the
decline rate. This may be significant in understanding the dramatic differences
for the TS model between the results of the MLCS2k2 fits and the SALT/SALT--II
fits, given that the latter rely on an empirical parameterisation in which
the effects of intrinsic colour dispersion are degenerate with those of
reddening by dust.

Much effort has gone into both theoretical and observational studies which
attempt to find direct correlations between \sn\ luminosity and particular
effects, including metallicity of the progenitor, age of the progenitor,
asymmetries of the explosion, central density and C/O ratio etc\footnote{For a
detailed list of references see Sec.~2 of \citet{Hoeflich10}.}. While these
effects could account for further secondary corrections to the light curve
fitting which need to be performed to account for intrinsic colour
variations, a great many more studies are required to sort out the physics.
Indeed, there is a possibility that different effects are involved in a
manner which may make them difficult to disentangle as the progenitor
population evolves over cosmological time scales.

It is well established that the age of progenitor system is a key variable
affecting SNeIa properties, a feature which has been known since the
early work of \citet{Hamuy95}, who observed that in their nearby sample,
brighter SNeIa tend to occur in late-type galaxies. A broad division of
\sne\ can be made into ``prompt'' and ``delayed'' groups
\citep{Scannapieco05}. The former group comprise intrinsically brighter
slow-declining \sne\ which come from a young stellar population and have
a rate proportional to the star formation rate ($\goesas0.5\,$Gyr timescale),
while the latter group consists of intrinsically dimmer fast decliners, which
take several Gyr to explode and come from much older populations with a rate
proportional to the mass of the host galaxy \citep{Sullivan06}. Since star
formation rate increases over the redshift ranges in which \sne\ occur, the
number of ``prompt'' \sne\ will increase with redshift, and therefore the
mean luminosity of the population should increase with redshift
\citep{Howell07}.

It is quite possible that it is intrinsic effects related
to the differences between the prompt and delayed events which are not
fully accounted for with the current light curve fitters. However, light
curve corrections can reverse the trends in the underlying population if
the fitter only assumes a single class of standardizable object, when there
is actually more than one. Recent studies by \citet{Sullivan10} and
\citet{Lampeitl10} both find a statistically significant correlation
between \sn\ luminosity and host galaxy type. In particular, more passively
evolving galaxies tend to host
\sne\ which, {\em after light curve correction} are of order $0.1\,$mag
brighter than those in galaxies with high specific star formation rates.
The passively evolving galaxies are generally more massive, and so there is
a related correlation, which has also been observed in earlier work with
smaller nearby samples \citep{Kelly10}.
\cite{Sullivan10} studied SNLS data using the SiFTO light curve fitter, which
uses a similar methodology to SALT. They found that events of the same
light-curve shape and colour were on average $0.08\,$mag brighter, at 4$\si$
confidence, for the the passively evolving subclass of \sne. The passively
evolving subclass also favoured smaller values of the SALT-like parameter
$\beta$ than those in galaxies with significant star formation, at the
$2.7\si$ confidence level. Very similar results were obtained by
\citet{Lampeitl10} using SNLS data and both the SALT-II and MLCS2k2 fitters.
For MLCS2k2, the favoured value of the parameter $\RV$ is found to be
different between the two subclasses, with values $\RV\goesas1$ favoured
by the passively evolving subclass and $\RV\goesas2$ by the the star-forming
hosts.

Both \citet{Sullivan10} and \citet{Lampeitl10} recommend correcting light
curves based on two sets of \sn\ templates, depending on galaxy types. Since
the effective \Omm\ parameter of the TS cosmology is more sensitive to
the differences between different light curve fitters, it would be extremely
interesting to test what effect this would have.

\section{Conclusion}

In conclusion, we find that the principal criticism of the TS cosmology
made by \citet{KFL} does not hold up when the issues surrounding the
systematics of \sn\ data reduction are thoroughly investigated. In particular,
the unreasonably small bestfit values of \Omm\ (or equivalently the
unreasonably high bestfit values of the void fraction, $\fvn$) for the
full Union and Constitution datasets are an artifact of failing to exclude
\sne\ below the scale of statistical homogeneity from the analysis. Such a
cut must be made for the purpose of consistency with the TS model, given
that an apparent Hubble bubble with certain characteristics will be found
below the SHS.
The main issue is not the size of the datasets, as \citet{KFL} claimed,
but the systematics of the data reduction methods.

We have shown that when suitable cuts are made then the
SALT/SALT-II fitters, as currently implemented, provide Bayesian evidence to
favour the spatially flat \LCDM\ model over the TS model. However, by
contrast the MLCS2k2 similarly provide Bayesian evidence that favours
the TS model over the spatially flat \LCDM\ model. Basically, both models
are a very good fit and it is the light curve fitting systematics that
underlie the few percent level differences which have to be sorted out to
distinguish the two cosmologies.

As yet these systematics are not fully understood, and involve many subtleties.
For example, the value of $\OmMn=0.42\pm0.10$ obtained
for the TS model with the SHS-cut SDSS-II sample is twice the
corresponding value for the SHS-cut MLCS17 sample,
$\OmMn=0.19^{+0.14}_{-0.17}$, despite their similar $\RV$ values. It
is clear that the differences do not involve a single parameter alone.
Nonetheless, given that an apparent Hubble bubble below
the scale of statistical homogeneity is a feature of the TS scenario,
the differences between SHS cuts applied to the MLCS17 and MLCS31 samples
suggest consistency for the TS scenario if galaxies which host ``normal''
\sne\ events have a reddening law with $\RV$ value close to the Milky way
value, $\RV\simeq3.1$, at least {\em at low redshifts}. Distinguishing
``normal'' \sne\ from other events may be further complicated by
\begin{itemize}
\item the existence of a subclass of events with nonstandard dust, possibly
related to circumstellar dust, as evidenced by the study of \citet{Wang09};
\item the existence of an intrinsic colour variation, uncorrelated with
decline rate, which distinguishes ``normal'' \sne\ in passively evolving
galaxies from those in galaxies with significant star formation, as evidenced
by many studies including the recent studies of \citet{Sullivan10} and
\citet{Lampeitl10}.
\end{itemize}

In our opinion systematic questions should ideally be resolved by detailed
studies which attempt to understand the astrophysics involved with as few
cosmological assumptions as possible, rather than purely empirical
correlations based on homogeneous cosmologies. For example, in trying to sort
out systematics at the percent level the current approach is often to test
how variation of empirical parameters affects Hubble residuals, using a
standard \LCDM\ model or a homogeneous dark energy cosmology with fixed
equation of state parameter, $w$. However, such a parameterisation has no
meaning for the TS cosmology, as was demonstrated in Fig.~3 of \citet{obs},
where the equivalent $w(z)$ determined from a perfectly smooth luminosity
distance relation was found to be ill--defined at $z\goesas1.7$.

Furthermore, from the viewpoint of the TS model distinctions based on
Hubble residuals should {\em not be used below the scale of statistical
homogeneity}, $z\lsim0.033$, since a natural variance in the Hubble
flow is to be expected below this scale. The only scales within which
Hubble residuals would be a safe determinant of empirical correlations
would be over those scales beyond the SHS over which an effective
linear global average Hubble law pertains, e.g., on scales $0.033\lsim
z\lsim0.1$. Beyond such a scale any Hubble residuals, whether based on
the \LCDM\ model, the TS model or any other model, are cosmology-dependent.

For a simple two model comparison it would be important to fully re-perform
the MLCS reduction, including cuts based on extinction priors, using the
TS model from the outset. Likewise, the changes to Hubble residuals with
different classes of \sn\ light curves for passively evolving galaxies,
as opposed to star-forming galaxies, should be investigated.

Given that the issue of dust extinction and reddening laws is so
entangled with the question of intrinsic colour variations, we really
require many independent studies, such as those of
\cite{Finkelman08,Finkelman10}, which examine reddening laws in other
galaxies without any reference to \sne.

Cosmology is a unique science, in the sense that its most basic quantity
-- distance -- can only be determined by assuming a cosmological model
when interpreting measurements such as spectra, apparent magnitudes and
angles on the sky. We have to be careful to recognize how the cosmological
models we assume affect our approach to data reduction.
The TS model is a well-motivated alternative to the standard \LCDM\ model
with some very different properties to homogeneous cosmologies and cannot
be parameterised by a well-defined dark energy equation of state parameter,
even though the luminosity distance is close to that of FLRW models.
The differences between the two models are at the same level as the
current systematic uncertainties in \sn\ data reduction which need to
be disentangled. We therefore hope that the TS model might be used alongside
the standard cosmology as a test bed for trying to determine which systematic
effects are truly astrophysical, and which might have an origin in
cosmological assumptions.

\section*{Acknowledgments}

This work was supported by the Marsden fund of the Royal Society
of New Zealand. We would like to thank Jay Gallagher for illuminating
discussions about reddening and extinction by dust, Alex Conley
for help with \cosfit, and the referee Martin Hendy for some constructive
comments.

\end{document}